\pdfoutput=1
\documentclass[11pt,twoside,a4paper,cmspaper,final,collab]{cms-tdr}
\def\svnVersion{d3eeea8}\def\svnDate{2026/02/02}\def\cmsCernNoTag{CERN-EP-2025-205}\def\cmsCernDate{\today}\def\cmsMessage{Published in Physical Review Letters as \href{http://dx.doi.org/10.1103/rk6w-1pcl}{\doi{10.1103/rk6w-1pcl}.}}
\begin{document}\cmsNoteHeader{TOP-24-009}

\newlength\cmsFigWidth
\newlength\cmsTabSkip\setlength{\cmsTabSkip}{2ex}

\ifthenelse{\boolean{cms@external}}{%
  \renewcommand{\appendixname}{End Matter}
  \newcommand{\addmatter}[1]{\appendixname~\ref{#1}\xspace}
  \providecommand{\cmsLeft}{upper\xspace}
  \providecommand{\cmsRight}{lower\xspace}
  \setlength{\cmsFigWidth}{0.45\textwidth}
}{
  \newcommand{\addmatter}[1]{\appendixname~\ref{#1}\xspace}
  \providecommand{\cmsLeft}{left\xspace}
  \providecommand{\cmsRight}{right\xspace}
  \setlength{\cmsFigWidth}{0.49\textwidth}
}

\newcommand{\pp}{{\Pp}{\Pp}\xspace}
\newcommand{\twz}{\ensuremath{\PQt\PW\PZ}\xspace}
\newcommand{\ttz}{{\ttbar}{\PZ}\xspace}
\newcommand{\ttw}{{\ttbar}{\PW}\xspace}
\newcommand{\fb}{\unit{fb}}
\newcommand{\DPhi}{\ensuremath{\Delta\phi}\xspace}
\newcommand{\DEta}{\ensuremath{\Delta\eta}\xspace}
\newcommand{\DeepJet}{\textsc{DeepJet}\xspace}
\newcommand{\MG}{\textsc{MG}\xspace}
\newcommand{\ParticleTransformer}{\textsc{ParticleTransformerAK4}\xspace}
\newcommand{\mZ}{\ensuremath{m_{\PZ}}\xspace}
\newcommand{\SRa}{\ensuremath{\text{SR}_{3\Pell}}\xspace}
\newcommand{\SRb}{\ensuremath{\text{SR}_{4\Pell}}\xspace}
\newcommand{\ZZ}{\ensuremath{\PZ\PZ}\xspace}
\newcommand{\WZ}{\ensuremath{\PW\PZ}\xspace}
\newcommand{\DY}{\ensuremath{\mathrm{DY}}\xspace} 
\newcommand{\CRZZ}{\ensuremath{\text{CR}_{\ZZ}}\xspace}
\newcommand{\CRWZ}{\ensuremath{\text{CR}_{\WZ}}\xspace}
\newcommand{\CRDY}{\ensuremath{\text{CR}_{\DY}}\xspace}
\newcommand{\CRTT}{\ensuremath{\text{CR}_{\ttbar}}\xspace}
\DeclareRobustCommand{\PV}{{\HepParticle{V}{}{}}\Xspace}
\DeclareRobustCommand{\PX}{{\HepParticle{X}{}{}}\Xspace}
\newcommand{\sobs}{5.8 }
\newcommand{\sexp}{3.5 }
\newcommand{\sobsTwo}{5.1 }
\newcommand{\sexpTwo}{3.0 }
\newcommand{\sobsThree}{3.4 }
\newcommand{\sexpThree}{2.1 }
\newcommand{\mutwz}{\ensuremath{\mu_{\twz}}\xspace}
\newcommand{\muttz}{\ensuremath{\mu_{\ttz}}\xspace}
\newcommand{\SSAll}{\ensuremath{1.77\pm0.32\,\text{(tot)}}\xspace}
\newcommand{\SSAllstatsyst}{\ensuremath{1.77\pm0.23\stat\pm0.22\syst}\xspace}
\newcommand{\SSTwo}{\ensuremath{1.82\pm0.38\,\text{(tot)}}\xspace}
\newcommand{\SSTwostatsyst}{\ensuremath{1.82\pm0.28\stat\pm0.26\syst}\xspace}
\newcommand{\SSThree}{\ensuremath{1.64\pm0.52\,\text{(tot)}}\xspace}
\newcommand{\SSThreestatsyst}{\ensuremath{1.64\pm0.42\stat\pm0.31\syst}\xspace}
\newcommand{\XSTwo}{\ensuremath{248\pm52\fb}\xspace} 
\newcommand{\XSTwostatsyst}{\ensuremath{248\pm 38\stat\pm35\syst \fb}\xspace}
\newcommand{\XSThree}{\ensuremath{242\pm77\fb}\xspace} 
\newcommand{\XSThreestatsyst}{\ensuremath{242\pm 62\stat\pm46\syst \fb}\xspace}
\newcommand{\SD}{s.d.\ }

\cmsNoteHeader{TOP-24-009}

\title{Observation of \texorpdfstring{\twz}{tWZ} production at the CMS experiment}

\abstract{
The first observation of single top quark production in association with a \PW and a \PZ boson in proton-proton collisions is reported. The analysis uses data at center-of-mass energies of 13 and 13.6\TeV recorded with the CMS detector at the CERN LHC, corresponding to a total integrated luminosity of 200\fbinv. Events with three or four charged leptons, which can be electrons or muons, are selected. Advanced machine-learning algorithms and improved reconstruction methods, compared to an earlier analysis, result in an unprecedented sensitivity to \twz production. The measured cross sections for \twz production are \XSTwo and \XSThree for $\sqrt{s}=13$ and 13.6\TeV, respectively. The signal is established with a statistical significance of \sobs standard deviations, with \sexp expected, compared to the background-only hypothesis. 
}

\hypersetup{
pdfauthor={CMS Collaboration},
pdftitle={Observation of tWZ production at the CMS experiment},
pdfsubject={CMS},
pdfkeywords={CMS, top quark, EW, cross section}}

\maketitle 

The standard model (SM) of particle physics provides a remarkably successful theoretical framework 
for describing the fundamental particles and their interactions through the electromagnetic, 
weak, and strong forces. Over the past several decades, the predictions of the SM have 
been extensively tested with high precision, and its theoretical consistency has withstood rigorous experimental scrutiny. 
Despite its successes, the SM is known to be an incomplete theory. For example, it does not account for 
dark matter~\cite{Bertone:2004pz}, the observed matter-antimatter asymmetry in the Universe~\cite{Canetti:2012zc}, 
neutrino masses~\cite{Deppisch:2015qwa}, and the quantum nature of gravity~\cite{Addazi:2021xuf}. 
Consequently, the search for phenomena not included in the SM remains a central focus of 
contemporary particle physics.

High sensitivity to deviations from the SM predictions, and therefore to beyond-the-SM (BSM) effects, 
can be achieved in direct searches~\cite{Rappoccio:2018qxp, ATLAS:2024itc, ATLAS:2024lda, ATLAS:2024fdw, CMS:2024phk, CMS:2024zqs, CMS:2024bni, CMS:2024zhe}, 
precision measurements~\cite{CMS:2024gzs}, 
studies of loop-induced processes~\cite{London:2021lfn}, 
or measurements of very rare processes. 
Among the rare processes predicted by the SM, the production of a single top quark \PQt in association 
with a \PW and a \PZ boson in proton-proton (\pp) collisions, $\pp\to\twz$, is particularly interesting~\cite{Maltoni:2019aot}.
This process is exceptionally rare with predicted cross sections of $136^{+9}_{-8}$\fb at $\sqrt{s}=13\TeV$ and 
$148^{+10}_{-9}$\fb at 13.6\TeV at next-to-leading order (NLO) 
accuracy in the strong coupling of quantum chromodynamics (QCD)~\cite{Alwall:2014hca, Frederix2012}. 
The \twz process provides direct access to the electroweak (EW) couplings of the top quark
through Feynman diagrams similar to the example shown in Fig.~\ref{fig:tWZ_feyn} (left). 
The \twz production allows for stringent tests 
of the gauge structure of the SM and the underlying symmetries governing 
particle interactions~\cite{Maltoni:2019aot, Faham:2021zet}. 
New heavy particles can induce anomalous top-quark couplings with the EW gauge bosons, 
as shown by the Feynman diagram in Fig.~\ref{fig:tWZ_feyn} (right). 
These would manifest as measurable deviations in the rate of \twz production, 
making measurements thereof a sensitive probe of BSM physics~\cite{Keaveney:2021dfa, Belvedere:2024wzg}. 
\ifthenelse{\boolean{cms@external}}
{
\begin{figure}[b]
\centering
\includegraphics[width=0.22\textwidth]{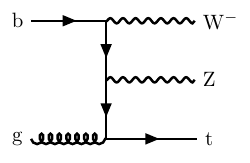}
\includegraphics[width=0.22\textwidth]{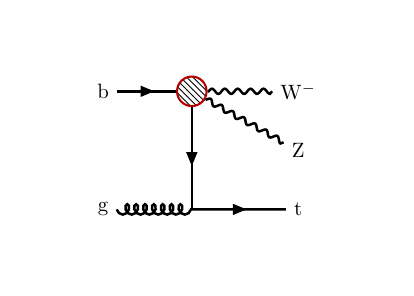}
\caption{Example Feynman diagrams for \twz production in the SM (left) and for anomalous \twz production 
through an effective interaction (right). Charge conjugate versions of these diagrams are implied.
  \label{fig:tWZ_feyn}
}
\end{figure}
}
{
\begin{figure}[b]
\centering
\includegraphics[width=0.32\textwidth]{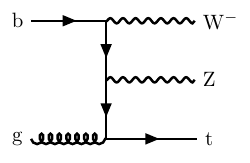}
\hspace{0.7cm}
\includegraphics[width=0.32\textwidth]{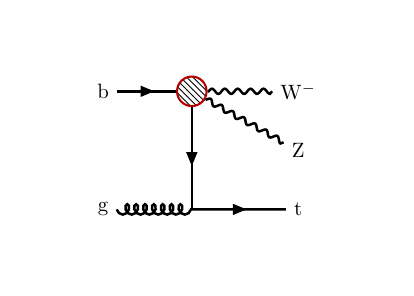}
\caption{Example Feynman diagrams for \twz production in the SM (\cmsLeft) and for anomalous \twz production 
through an effective interaction (\cmsRight). Charge conjugate versions of these diagrams are implied.
  \label{fig:tWZ_feyn}
}
\end{figure}
}

This Letter presents an analysis of \pp collision data from the CERN LHC, aiming to identify events 
from \twz production, characterized by 
the presence of multiple leptons, jets, and missing transverse momentum (\ptmiss). 
However, the experimental challenges are significant. 
The small production cross section of the \twz process compared to backgrounds with a similar signature, 
namely \ttz and diboson production with additional jets, necessitates advanced analysis techniques to isolate the signal. 
Evidence for \twz production has been reported by the CMS Collaboration in 2024 using 
$\sqrt{s}=13\TeV$ data corresponding to an integrated luminosity of 138\fbinv, recorded in 2016--2018~\cite{twzevidence}. 
The previous analysis reported an observed statistical significance of 3.4 standard deviations (\SD\unskip) with 
1.4 \SD expected. 
The analysis presented in this Letter uses state-of-the-art machine-learning (ML) algorithms 
and improved reconstruction methods to study \twz production with unprecedented sensitivity. 
It updates the analysis of 13\TeV data and includes, in addition, 
\pp collision data at 13.6\TeV recorded in 2022--2023. The total integrated luminosity of the dataset is 
200\fbinv, of which 62\fbinv were recorded at 13.6\TeV.
Tabulated results are provided in the HEPData record for this analysis~\cite{hepdata}.

The CMS apparatus~\cite{CMS:2008xjf,CMS:2023gfb} is a multipurpose, nearly hermetic detector, 
designed to trigger on~\cite{CMS:2020cmk,CMS:2016ngn,CMS:2024aqx} and identify electrons, muons, photons, 
and hadrons~\cite{CMS:2020uim,CMS:2018rym,CMS:2014pgm}. A global ``particle-flow'' (PF) algorithm~\cite{CMS:2017yfk} 
aims to reconstruct all individual particles in an event, combining information provided by the 
all-silicon inner tracker and by the crystal electromagnetic (ECAL) and brass-scintillator hadron calorimeters (HCAL), 
operating inside a 3.8\unit{T} superconducting solenoid, with data from the gas-ionization muon detectors 
embedded in the flux-return yoke outside the solenoid. 
The reconstructed particles are used to build \PGt lepton candidates, jets, and \ptmiss~\cite{CMS:2018jrd,CMS:2016lmd,CMS:2019ctu}. 

Events of interest are selected using a two-tiered trigger system. The first level, composed 
of custom hardware processors, uses information from the calorimeters and muon detectors to select 
events at a rate of around 100\unit{kHz} within a fixed latency of about 4\mus~\cite{CMS:2020cmk}. 
The second level, known as the high-level trigger, consists of a farm of processors running a 
version of the full event reconstruction software optimized for fast processing, and reduces the event 
rate to a few kHz before data storage~\cite{CMS:2016ngn,CMS:2024aqx}.
Events are selected using a combination of triggers that require the presence of one, two, or three 
reconstructed leptons. The trigger efficiency is close to 100\% for signal events passing the analysis 
selection~\cite{CMS:2021ugv}.

The simulation of the signal is carried out using the Monte-Carlo (MC) event generator  
\MGvATNLO (\MG) version 2.6.5~\cite{Alwall:2014hca, Frederix2012} at NLO accuracy in perturbative QCD and in 
the five-flavor scheme (5FS). The factorization and renormalization scales are set 
to the default dynamical scales in \MG, which are calculated as half the sum of all final state
transverse energies~\cite{Hirschi:2015iia}. 
The subsequent decays of the \PQt quarks, \PW and \PZ bosons are simulated using \textsc{MadSpin}~\cite{Frixione:2007zp, Artoisenet:2012st}.
At leading order (LO) accuracy and in the 5FS, \twz can be identified through the 
tree-level partonic process $\Pg \PQb \to \twz$. 
However, at NLO, real corrections of the type $\Pg \Pg \to \PQt \PW \PZ \PQb$ arise, that can 
feature a resonant top quark pair (\ttbar) in the intermediate state and therefore cause overlap 
with the LO \ttz process~\cite{Faham:2021zet}. In other words, the emission of an extra 
\PQb quark implies that the $\PQt \PW \PZ \PQb$ final state can also occur from the process 
$\Pg \Pg \to \ttz$, with $\PQt \to \PQb\PWp$ and $\PAQt \to \PAQb \PWm$, namely from LO \ttz production. 
The overlap removal is performed using the \textsc{MadSTR} plugin~\cite{Frixione:2019fxg}, 
which implements both diagram removal (DR) and diagram subtraction (DS) 
techniques~\cite{Frixione:2008yi, White:2009yt}. 
For the signal samples, two DR options are used: DR1, which removes from the process amplitude 
both the squared resonant term and the interference term between resonant and nonresonant contributions, 
and DR2, which removes only the resonant term. The nominal samples are generated using the DR1 scheme. 

The main background processes are simulated with different MC event generators. 
The dominant backgrounds are generated at NLO accuracy in perturbative QCD, 
where \ttz and $\PQt \PZ \Pq$ are simulated using \MG, while $\PW \PZ$ production is 
simulated using \POWHEG~v2~\cite{Nason:2004rx,Frixione:2007vw,Frixione:2007nw,Alioli:2010xd,Re:2010bp}.
Because of larger uncertainties in $\PW \PZ$ production in association with a \PQb quark 
compared to jets from light-flavor quarks and gluons, we split this background into \WZ{}+\PQb and 
\WZ{}+j, respectively. 
Smaller backgrounds are simulated using \MG, \POWHEG and \MCFM~v7.0.1~\cite{Campbell:1999ah, Campbell:2011bn, Campbell:2015qma}. 
These comprise di- and tri-boson processes, consolidated under $\PV\PV(\PV)$ where 
$\PV = \PW$ or \PZ, additional processes with a photon in the final state, $\PX\PGg$, 
and processes in which a single top quark or a \ttbar 
pair is produced in association with additional particles, $\PQt(\PAQt)\PX$, 
grouping $\PQt\PW$, $\ttbar\PH$, $\ttbar\ttbar$, 
$\ttbar\PW\PW$, $\ttbar\PW\PZ$, and $\ttbar\PZ\PZ$.
The NNPDF~3.1 next-to-NLO (NNLO) parton distribution functions (PDFs)~\cite{NNPDF:2017mvq} 
are used for the generation of all samples, 
together with \PYTHIA~8.240~\cite{Sjostrand:2014zea} for the modeling of the parton shower (PS) and hadronization. 
The \ttz background cross section 
is scaled to account for approximate next-to-NNLO (a$\text{N}^3$LO) QCD and NLO EW 
corrections~\cite{Kidonakis:2024lht}. 
The underlying event is described using the CP5 tune~\cite{TuneCP5}. 
Matrix element calculations and PS simulations are matched using the
FxFx~\cite{Frederix2012} and MLM~\cite{Alwall:2007fs} algorithms for processes simulated with 
\MG at NLO and LO, respectively.
The detector response is simulated with a detailed description of the CMS apparatus based on the 
\GEANTfour toolkit~\cite{Agostinelli:2002hh}. Additional \pp interactions occurring in the same or 
adjacent bunch crossings (pileup) are modeled using \PYTHIA.
All simulated samples are produced taking into account the LHC and detector conditions 
of the different data-taking periods.

The PF algorithm~\cite{CMS:2017yfk} is used to reconstruct and identify 
individual particles in the event, combining information from different parts of the detector. 
The primary vertex (PV) is taken to be the vertex corresponding to the hardest scattering in the event, 
identified using the tracking information alone, as described in Section 9.4.1 of Ref.~\cite{CMS-TDR-15-02}. 
The energy of electrons is determined from a combination of the electron momentum 
from the reconstructed track,
the energy of the corresponding ECAL cluster, and the energy sum of all bremsstrahlung photons spatially 
compatible with originating from the electron track. The momentum of muons is obtained from 
the curvature of their tracks measured in the tracking system and in the muon detectors. 
The energy of charged hadrons is evaluated via a combination of their momentum measured 
in the tracker and the matching of the ECAL and HCAL energy deposits, corrected for the 
response function of the calorimeters to hadronic showers. Finally, the energy of neutral hadrons is 
obtained from the corresponding corrected ECAL and HCAL energies.

Electrons and muons must have a transverse momentum $\pt > 10\GeV$ for events
collected at $\sqrt{s}=13\TeV$, and $\pt > 15\GeV$ for events collected at $\sqrt{s}=13.6\TeV$. 
In addition, electrons must have pseudorapidity $\abs{\eta} < 2.5$ and muons $\abs{\eta} < 2.4$. 
In this Letter, nonprompt leptons denote reconstructed leptons not originating from 
the decay of a \PW or \PZ boson. 
To reduce their presence in the final selection, multivariate discriminants are employed~\cite{CMS:2023dsu}, 
obtained from dedicated trainings on 13\TeV~\cite{CMS:2023ftu} and 13.6\TeV~\cite{CMS:2024ild} data. 
The lepton identification algorithms for 13\TeV data have efficiencies of 87 and 92\% 
for electrons and muons, respectively. 
An improved identification is used for 13.6\TeV data, 
resulting in lower efficiencies of 60 and 84\% for electrons and muons, 
but with a threefold improvement in the rejection of nonprompt leptons. 

Jets are reconstructed by clustering the PF candidates using the anti-\kt 
algorithm~\cite{Cacciari:2008gp, Cacciari:2011ma} with a distance parameter of 0.4. 
Jets are required to have $\pt > 25\GeV$, $\abs{\eta} < 2.5$, and to be separated from 
selected leptons by a distance $\DR = \sqrt{\smash[b]{(\DEta)^2 + (\DPhi)^2}} > 0.4$. 
Here, \DEta and \DPhi are the differences in pseudorapidity and azimuthal angle between the jet and lepton, respectively. 
Jets are corrected for pileup contributions and detector effects as 
detailed in \addmatter{endmatter:pileup}.
The \DeepJet algorithm~\cite{Sirunyan:2017ezt, Bols:2020bkb, CMS:DP-2023-005} is used in 13\TeV data 
to identify jets that originate from the fragmentation of \PQb quarks (\PQb jets). 
For 13.6\TeV data, an improved algorithm~\cite{Qu:2022mxj, CMS-DP-2022-050, CMS-DP-2022-049} 
is employed with better performance~\cite{CMS-DP-2024-024, CMS-DP-2024-025}. 
We define \PQb jets to be jets that pass a requirement on the algorithm score with  
an identification efficiency of around 80\% for jets truly originating from \PQb quarks
and misidentification probabilities of about 1\% for light-quark and gluon jets and 
about 14\% for charm quark jets.

We require events in the final selection to have at least three reconstructed leptons, the \pt-leading 
lepton must have $\pt>25\GeV$ and the \pt-subleading one must have $\pt>15\GeV$.
The events are split into different regions according to the number of leptons, 
jets, \PQb jets, \ptmiss, and invariant mass of the lepton pairs with opposite sign and same 
flavor (OSSF). The OSSF lepton pairs are considered to lie inside the \PZ boson mass window 
if the invariant mass is within $\mZ \pm 15 \GeV$, where $\mZ=91.2\GeV$~\cite{ParticleDataGroup:2024cfk}. 
The signal region (SR) is divided into two parts, depending on the number of leptons: 
events with three leptons enter the \SRa, while events with four leptons are assigned to the \SRb. 
In the \SRa, one OSSF lepton pair must be present with mass within the \mZ window. 
The events must also contain two jets, at least one of which is \PQb tagged. 
In the \SRb, selected events must have exactly one OSSF lepton pair within 
the \mZ window. Additionally, at least one \PQb-tagged jet must be present in the event. 

In addition to the two SRs, we define four control regions (CRs) to constrain 
the most important backgrounds and validate their modeling. 
Two CRs are constructed to constrain the \ZZ and \WZ backgrounds. 
The other two help to improve the modeling of events containing nonprompt leptons, 
originating from Drell--Yan (\DY) and \ttbar production. 
Events containing at least one jet and exactly four leptons with two OSSF lepton pairs 
inside the \mZ window enter the \CRZZ, which is enriched in events from \ZZ production. 
The \CRWZ contains events with at least two jets, 
no \PQb jets, three leptons, two of which are within the \mZ window, and $\ptmiss > 40\GeV$. 
The \CRWZ is also used to improve the description of the jet multiplicity for the \WZ process, as
detailed in the \addmatter{endmatter:njet}.
Events entering the \CRDY satisfy the same requirements as for the \CRWZ, but with an inverted 
\ptmiss selection and no requirement on the number of jets. 
Finally, the \CRTT contains events with at least two jets, one of which is \PQb tagged, 
three leptons but no OSSF lepton pair inside the \mZ window.

The contribution of the nonprompt background is estimated from a fit to data
simultaneously with the signal extraction. In the fit, the misidentification rates 
are left as free parameters and are constrained with the help of 
distributions in the trailing lepton \pt measured in the \CRDY and \CRTT. 
The trailing lepton is defined as the selected lepton with the smallest \pt.
Additional details on the estimation of the nonprompt-lepton background are provided in \addmatter{endmatter:nonprompt}.

A transformer ML algorithm based on the particle transformer \textsc{ParT} 
architecture~\cite{Qu:2022mxj} is employed to discriminate between signal and 
background processes, mainly \ttz and \WZ production. 
Four different dedicated models are trained: two for \SRa and two for \SRb, 
corresponding to trainings with 13 and 13.6\TeV data, respectively. 
In \SRa, the \textsc{ParT} algorithm is set up to have four output nodes, each representing 
the classification as \twz, \ttz, \WZ, or any of the minor remaining backgrounds. 
The events are then divided into three categories based on the maximum value of the output nodes: 
\twz, \ttz, and other backgrounds which combine \WZ and minor backgrounds. 
This results in a region enriched in signal events, 
a second region mainly used to constrain the \ttz background and related uncertainties, 
and a third region used to constrain other backgrounds. 
In \SRb, a binary classifier to distinguish \twz from background is used 
due to the lower statistical precision of the data events with four leptons. 
More details on the algorithm architecture and input variables are provided in \addmatter{endmatter:part}.

The \twz cross section is measured by maximizing a binned likelihood function in a fit to the data. 
This function combines the Poisson probabilities of the yields with the predicted signal 
and background estimates in each bin. It also includes all sources of systematic uncertainties, 
represented as nuisance parameters, which may influence the observed number of signal or background events. 
The fit yields the measured signal strength, defined as the ratio of the observed cross section to the 
expected value, along with its uncertainty. The fit is performed using the CMS statistical analysis 
program \textsc{Combine}~\cite{CMS:2024onh}. Different distributions enter the fit according to the 
SR or CR, separately for each center-of-mass energy. 
Three distributions in the classifier output nodes with five bins each are measured in the \SRa, 
corresponding to the values of the \twz, \ttz, and other background nodes. 
Two bins from the classifier output score in the \SRb enter the fit. 
The total number of events measured in the \CRZZ enters the fit to constrain the normalization of the \ZZ background.  
The jet multiplicity distribution is measured in the \CRWZ to constrain the uncertainty on the modeling of 
additional jets produced in association with \WZ. The distribution of the trailing lepton \pt 
is used for the \CRDY and \CRTT to constrain the rates of the nonprompt-lepton identification. 

The distributions in the \twz and \ttz output nodes in \SRa, along with the output of the classifier in \SRb, 
are shown in Fig.~\ref{fig:MLoutput}. 
The distributions are shown for the best fit values of the background normalizations and nuisance parameters 
corresponding to systematic uncertainties, after a fit to the data. 
\ifthenelse{\boolean{cms@external}}
{
\begin{figure}[tbp]
\centering
\includegraphics[width=0.35\textwidth]{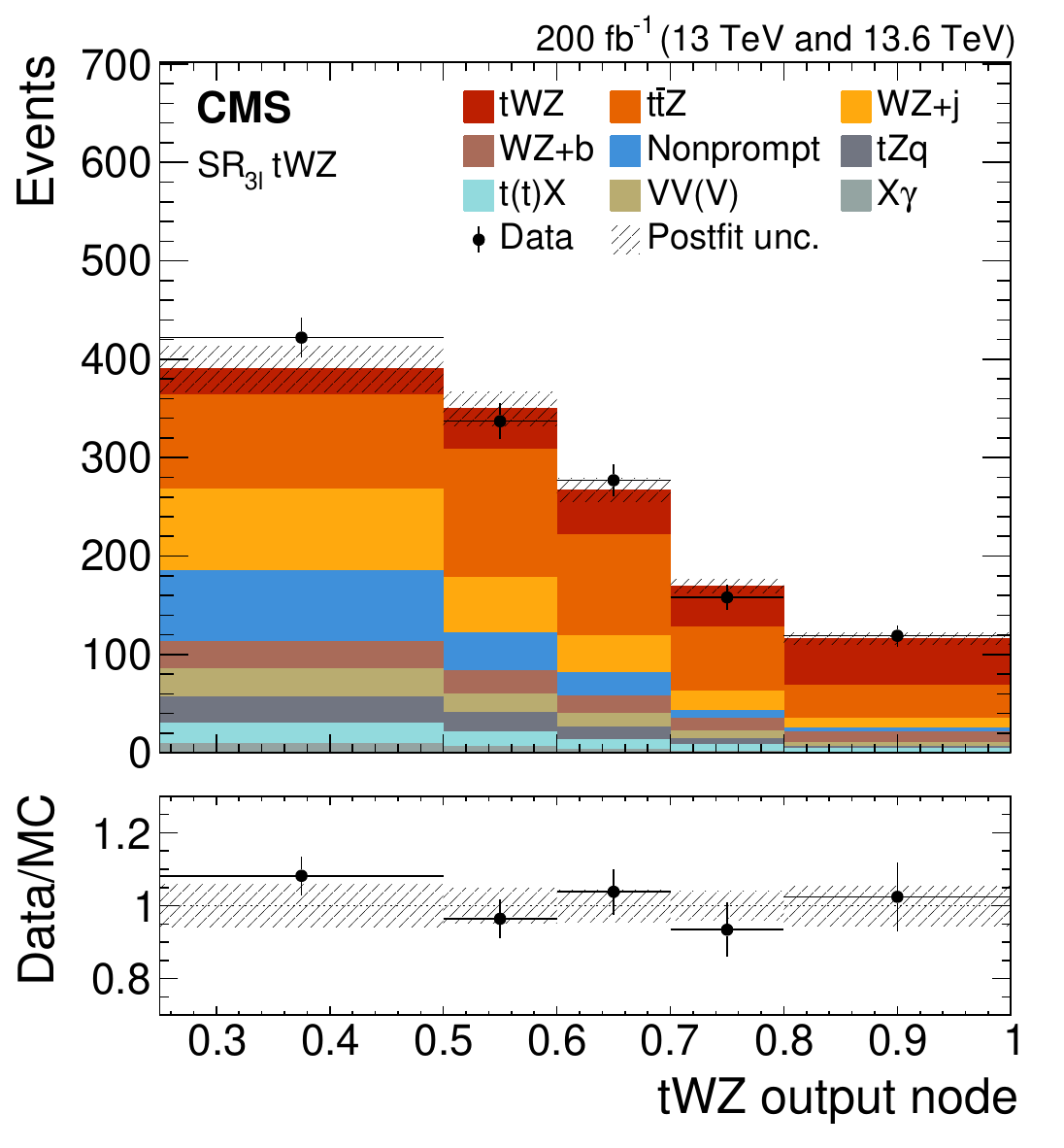}
\includegraphics[width=0.35\textwidth]{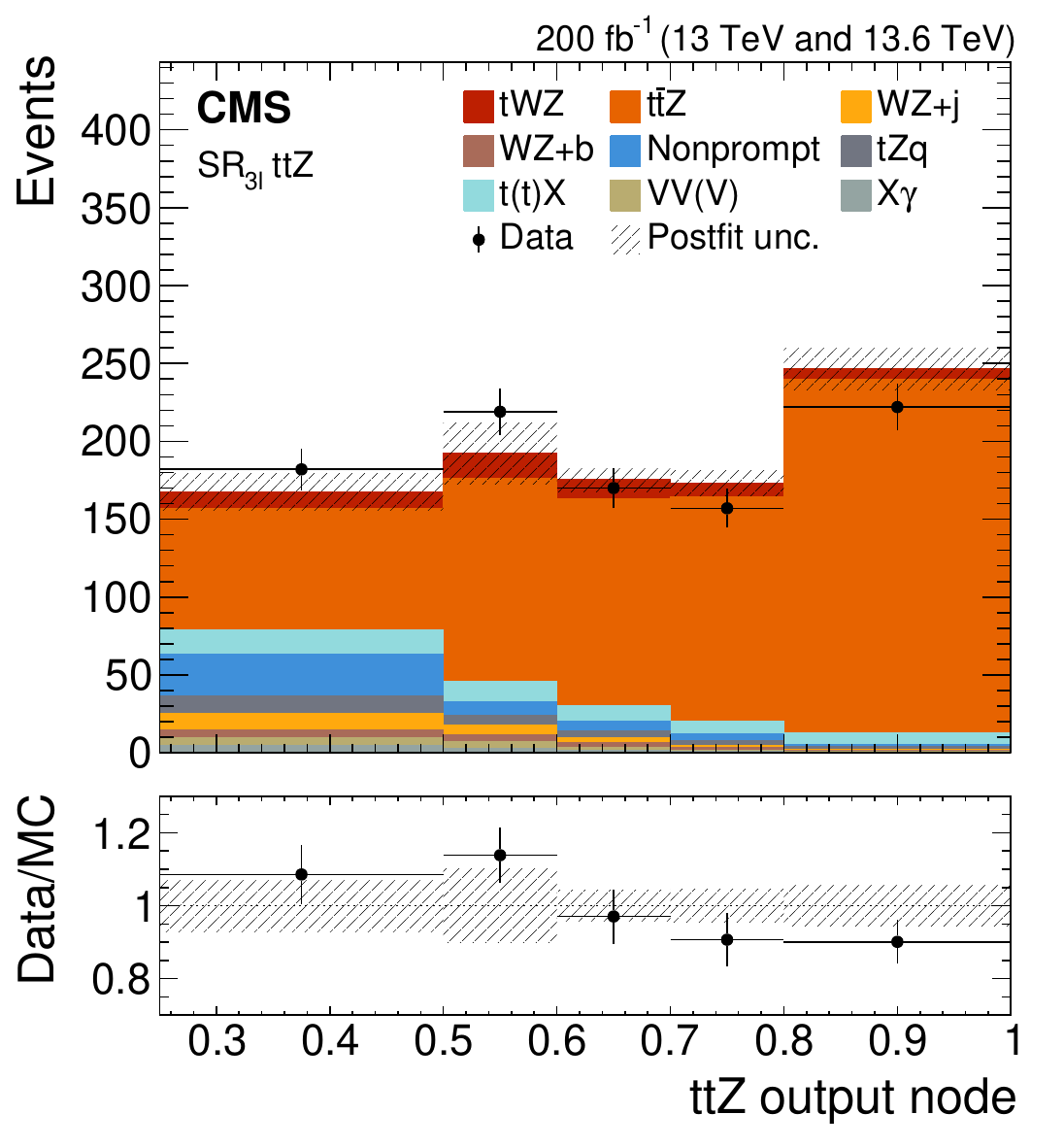}
\includegraphics[width=0.35\textwidth]{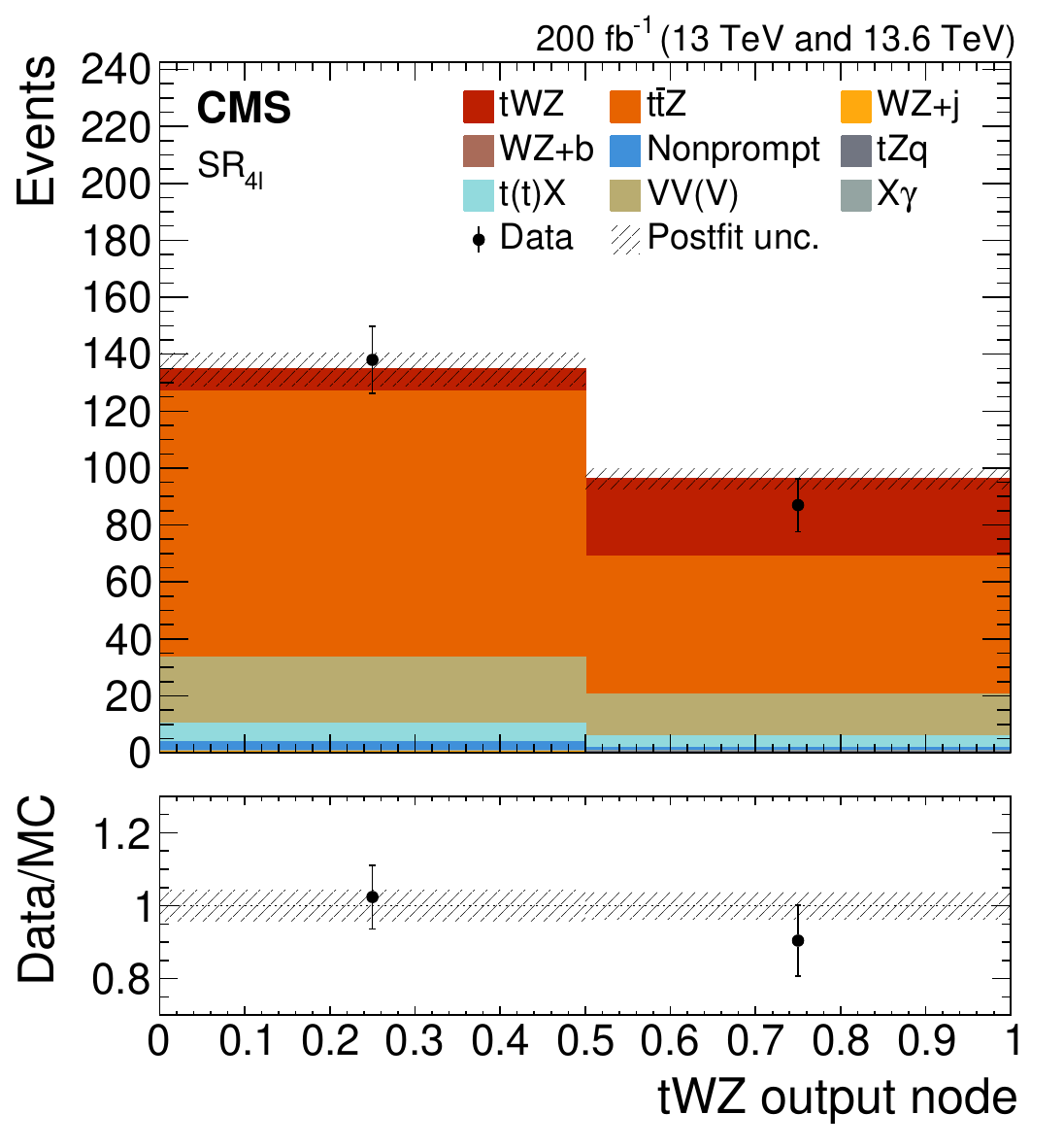}
\ifthenelse{\boolean{cms@external}}
{
\caption{Distributions in the classifier output nodes for the \twz (upper) and \ttz (middle) 
nodes in \SRa, and in the classifier output node in \SRb (lower) after a fit to the data. 
The dashed band shows the postfit uncertainty. 
The results from fits to 13 and 13.6\TeV data are combined for illustration purposes. 
The ratio of data to the predictions from MC simulation is shown below the distributions. 
\label{fig:MLoutput}
}
}
{
\caption{Distributions in the classifier output nodes for the \twz (upper left) and \ttz (upper right) 
nodes in \SRa, and in the classifier output node in \SRb (lower) after a fit to the data. 
The dashed band shows the postfit uncertainty. 
The results from fits to 13 and 13.6\TeV data are combined for illustration purposes. 
The ratio of data to the predictions from MC simulation is shown below the distributions. 
\label{fig:MLoutput}  
}
}
\end{figure}
}
{
\begin{figure}[tbp]
\centering
\includegraphics[width=\cmsFigWidth]{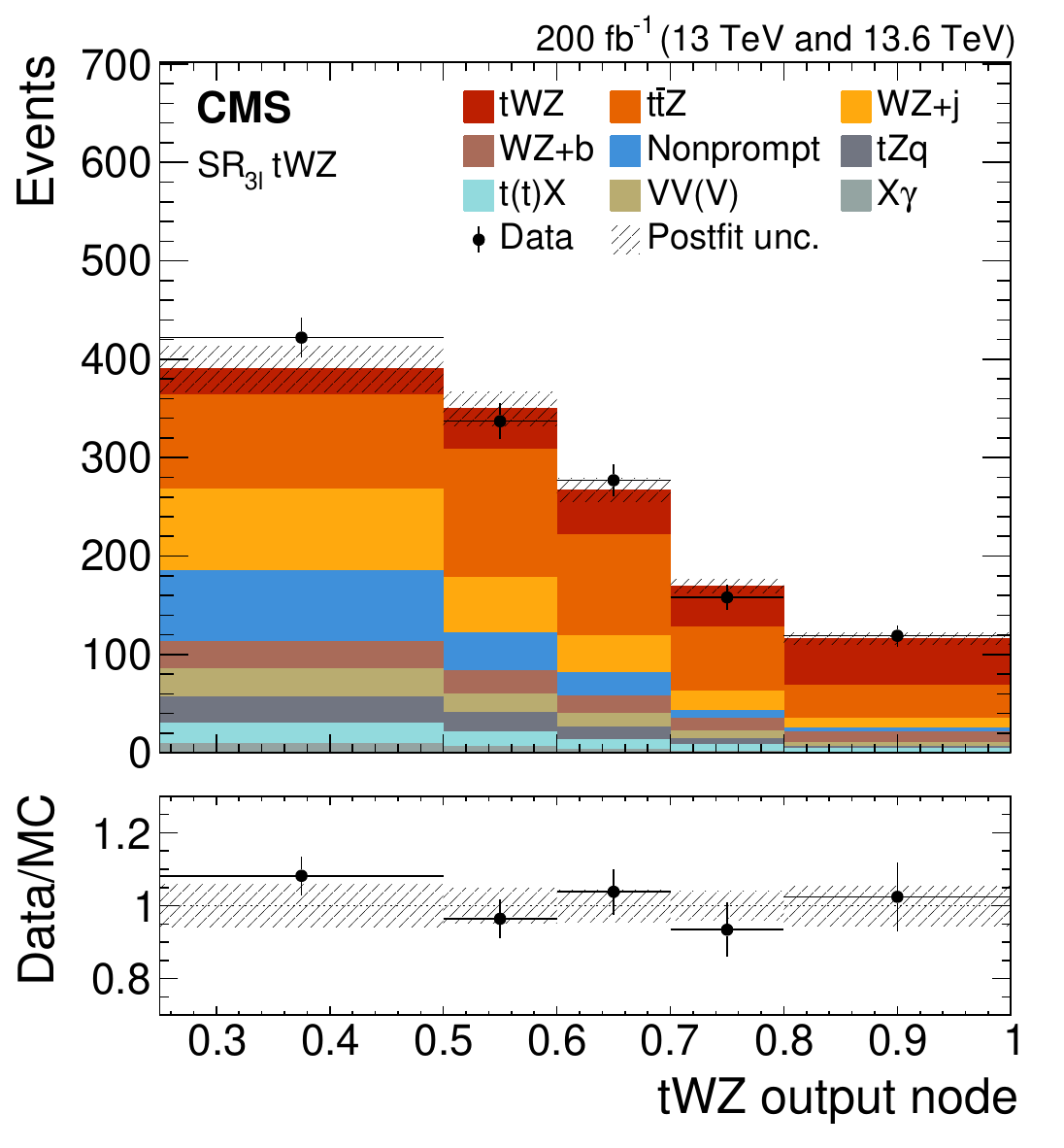}
\includegraphics[width=\cmsFigWidth]{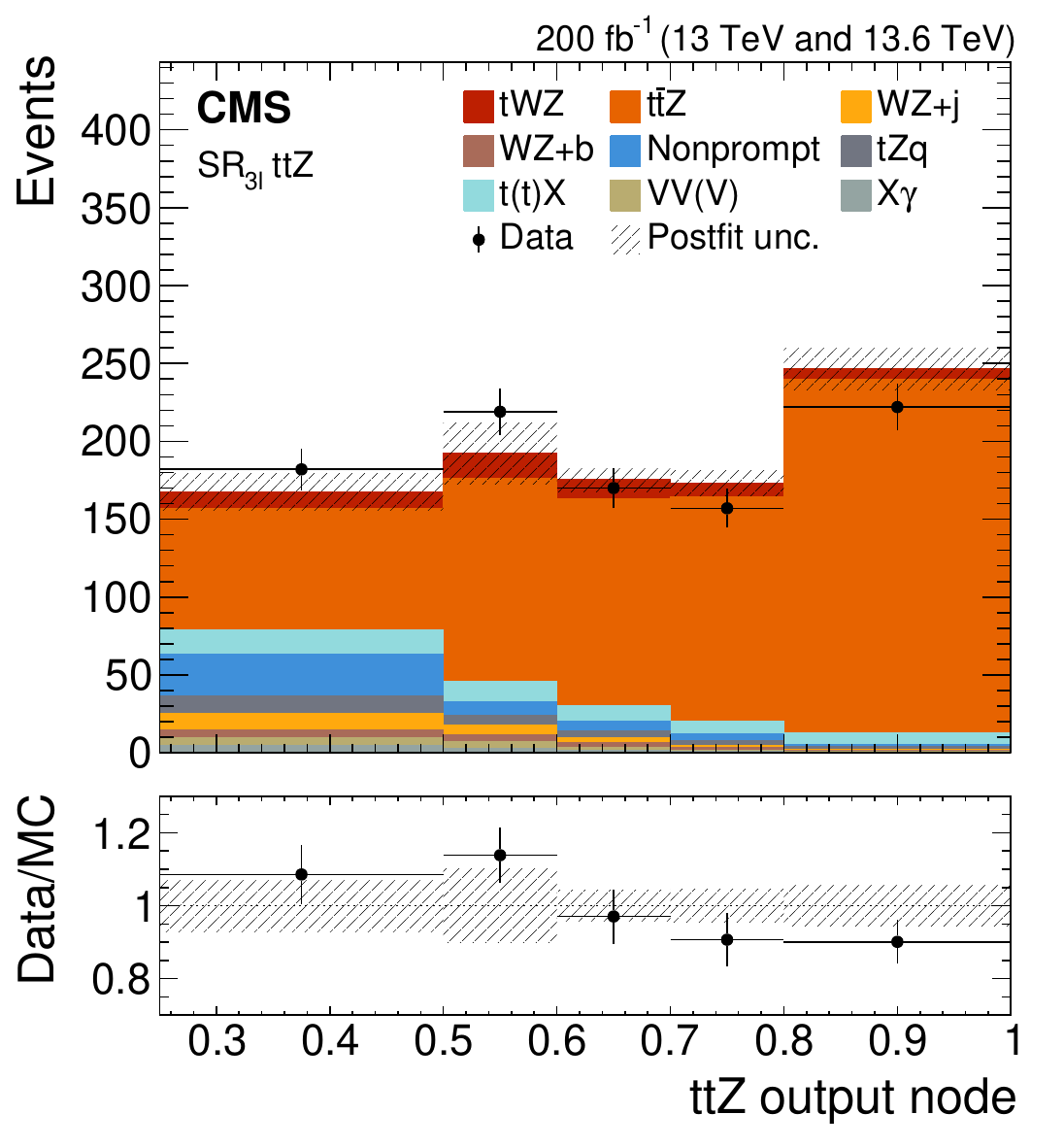}
\includegraphics[width=\cmsFigWidth]{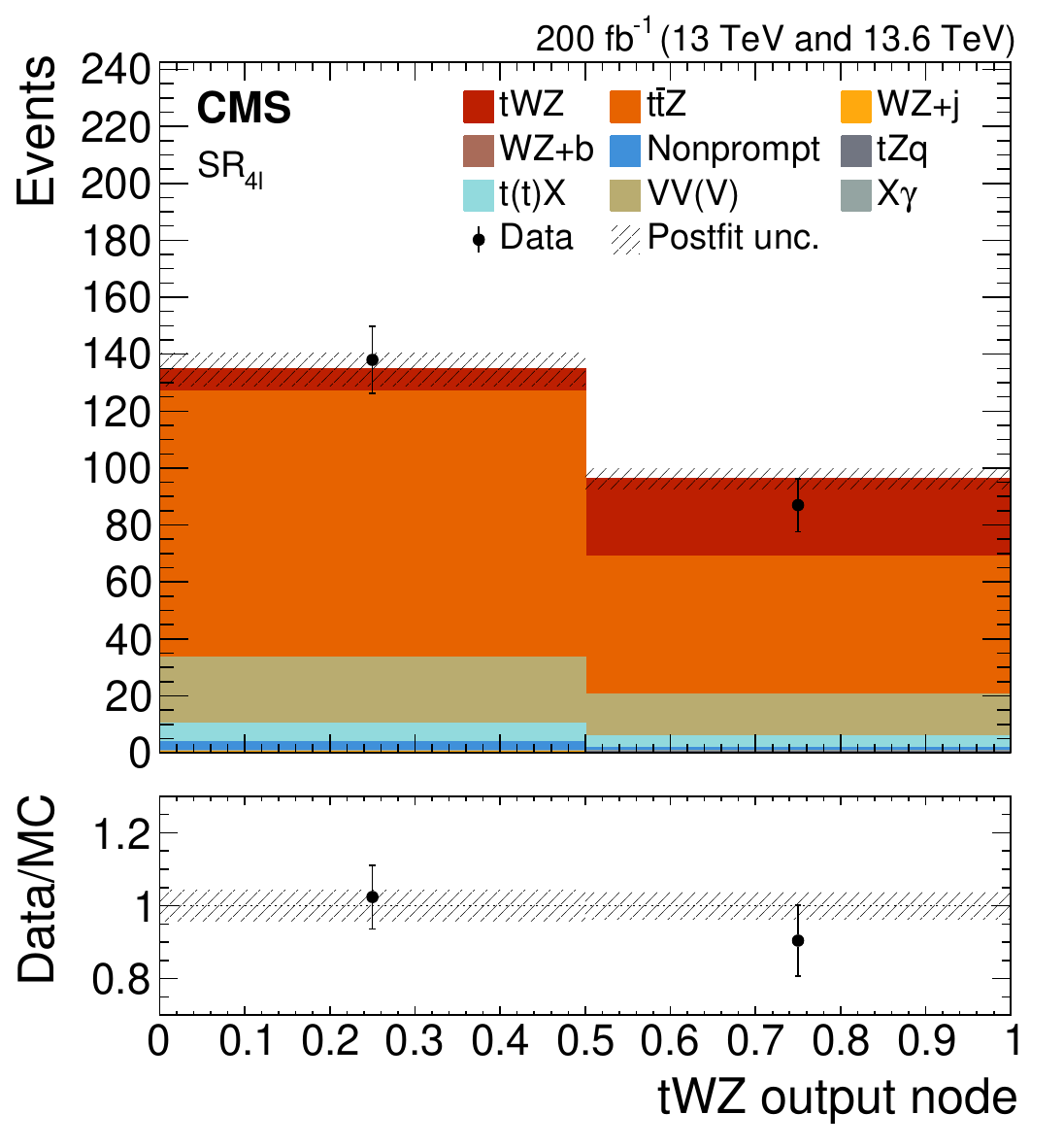}
\caption{Distributions in the classifier output nodes for the \twz (upper \cmsLeft) and \ttz (upper \cmsRight) 
nodes in \SRa, and in the classifier output node in \SRb (lower) after a fit to the data. 
The dashed band shows the postfit uncertainty. 
The results from fits to 13 and 13.6\TeV data are combined for illustration purposes. 
The ratio of data to the predictions from MC simulation is shown below the distributions. 
\label{fig:MLoutput}
}
\end{figure}
}

Several sources of systematic uncertainties are considered, comprising experimental and theoretical components. 
The experimental ones are the uncertainties in the jet energy corrections and resolution~\cite{CMS:2016lmd}, 
unclustered energy~\cite{CMS:2019ctu}, the description of pileup in the simulation~\cite{CMS:2020ebo}, 
the trigger~\cite{CMS:2016ngn, CMS:2020cmk}, \PQb tagging~\cite{Sirunyan:2017ezt}, 
and lepton efficiency corrections~\cite{CMS:2018rym, CMS:2020uim}. 
The integrated luminosity has an uncertainty of 1.6\% for the 2016--2018 
data-taking years~\cite{CMS-LUM-17-003,CMS-PAS-LUM-17-004,CMS-PAS-LUM-18-002}, 
and 1.4 and 1.3\% for 2022 and 2023, respectively~\cite{CMS-PAS-LUM-22-001, CMS-DP-2024-068}. 
The theoretical uncertainties in the signal and main backgrounds include 
uncertainties from the PDFs, the matrix element renormalization and factorization scales, 
and the PS simulation, namely initial- and final-state radiation and the 
matching of the PS to the matrix-element calculations.
Additional sources of uncertainty affecting the signal are taken into account by considering 
differences between the DR1 and DR2 subtraction schemes. 
Variations from DS are not considered, as these are smaller than the ones derived with DR2. 
The uncertainty associated with the modeling of additional jets in \WZ production is also included, more details 
are given in the \addmatter{endmatter:njet}. 
In addition, theoretical uncertainties in the cross sections of the background processes are considered. 
A normalization uncertainty of 7\% is employed for the \ttz cross section~\cite{CMS:2024mke, ATLAS:2023eld}. 
Two different normalization uncertainties are considered for the \WZ and \ZZ processes, 
where we assign a 20\% uncertainty on \PV{}\PV{}+\PQb production~\cite{CMS:2021ugv} 
and a 10\% uncertainty on \PV{}\PV{}+j~\cite{CMS:2021icx}. 
An uncertainty of 11\% is used for $\PQt \PZ \PQq$ production~\cite{CMS:2021ugv}, 
and a 10\% uncertainty is used for $\PV\PV(\PV)$~\cite{CMS-SMP-19-014, ATLAS-WGG} 
and $\PX\PGg$~\cite{PhysRevLett.132.021802, CMS-SMP-22-006}. 
Uncertainties of 5 and 10\% are applied to the \ttbar~\cite{Czakon:2011xx} and \DY processes~\cite{CMS:2022ilp}, respectively.  
A 25\% uncertainty is considered for the \ttw process where the measurements 
do not agree with the SM predictions by about this amount~\cite{CMS:2022tkv}, 
and a 20\% uncertainty is assigned to $\PQt(\PAQt)\PX$~\cite{CMS:2022ytw, CMS:2022dwd, CMS:2023ftu}. 
Additional uncertainties describe the rate of misidentified nonprompt leptons as detailed in \addmatter{endmatter:nonprompt}. 
Finally, uncertainties arising from the finite size of simulated samples are accounted for using 
the Barlow--Beeston light method~\cite{Barlow:1993dm}.
We treat experimental uncertainties and the rate of misidentified nonprompt leptons 
as uncorrelated between the 13 and 13.6\TeV datasets, 
while the theoretical uncertainties and background normalizations are fully correlated.

\begin{figure}[tbp]
\centering
\includegraphics[width=\cmsFigWidth]{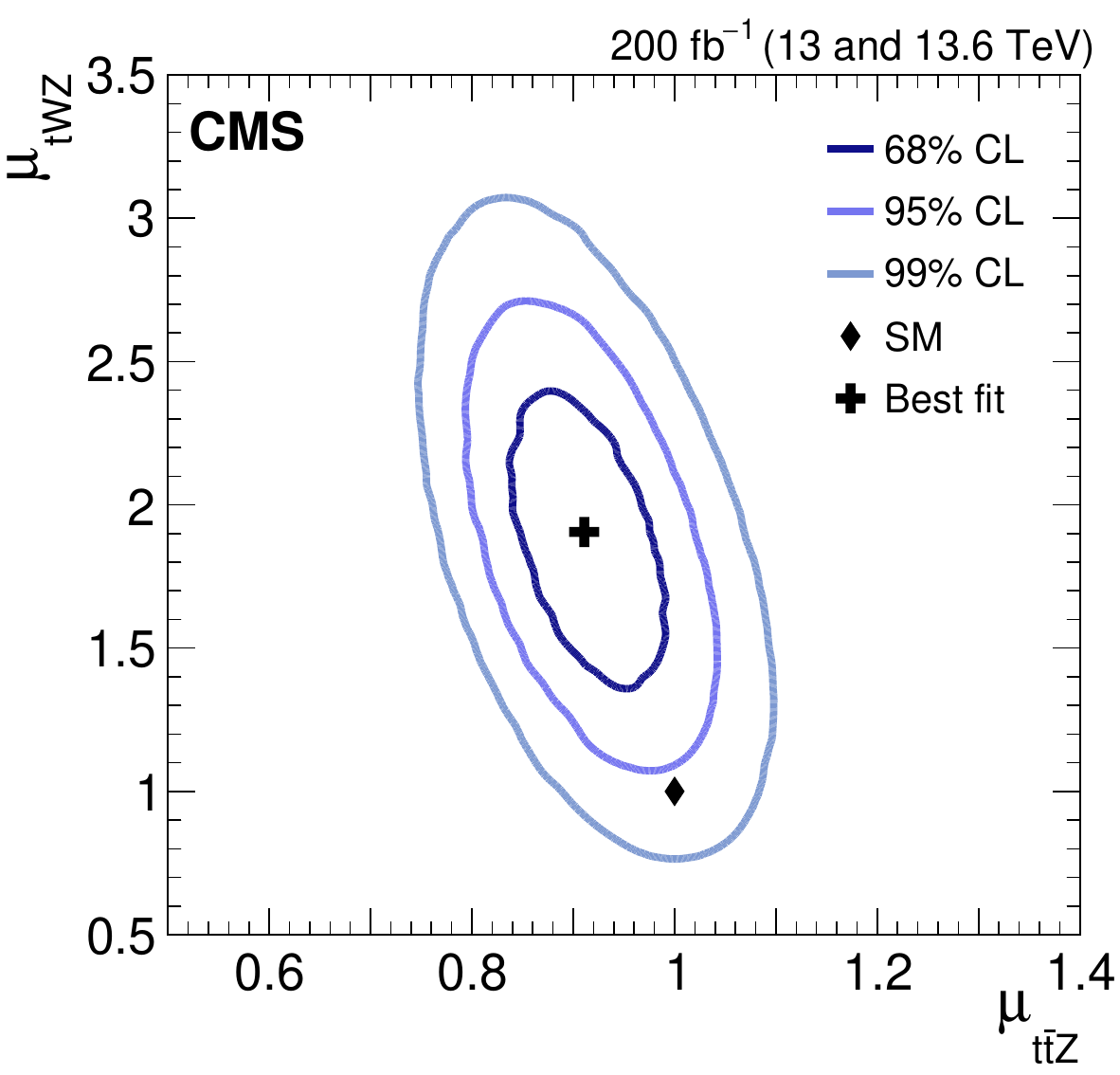}
\caption{Likelihood scan of the signal strengths for \twz and \ttz production. 
The black cross shows the best fit value, while the black diamond indicates the SM expected value. 
The contours in different shades of blue correspond to the 68, 95, and 99\% confidence levels.
\label{fig:simultaneousFit}
}
\end{figure}

The measured signal strength of \twz production is $\mutwz=\SSAllstatsyst$ with an observed 
statistical significance of \sobs \SD compared to the background-only hypothesis, with \sexp \SD expected. 
The theoretical predictions come with an uncertainty on the signal strength of 0.06, 
thus the result is 2.3 \SD larger than the SM expectation, 
consistent with the previous measurement~\cite{twzevidence}.
The signal strength using 13\TeV data only is \SSTwostatsyst, in agreement with the one 
from 13.6\TeV data only, \SSThreestatsyst. 
The corresponding observed significances are \sobsTwo \SD and  
\sobsThree \SD for 13 and 13.6\TeV data, respectively, with expected significances 
of \sexpTwo \SD and \sexpThree \SD\unskip, respectively.
The measured cross sections for \twz production are \XSTwostatsyst and \XSThreestatsyst 
for $\sqrt{s}=13$ and 13.6\TeV, respectively. 
Signal strengths measured in the different data-taking periods and leptonic channels 
are presented in \addmatter{endmatter:comp}. 
The dominant systematic uncertainty is the normalization of the \ttz process 
with a contribution of 14\% to the total uncertainty. Other relevant uncertainties are the 
normalization of the \WZ{}+\PQb process and the uncertainties 
in the jet energy corrections, with contributions of 9 and 6\%, respectively. 
We determine the signal strengths of \twz and \ttz production in a simultaneous fit, 
without external constraints on the \ttz cross section. The result of this fit is shown 
in Fig.~\ref{fig:simultaneousFit}.
The measured signal strengths are $\mutwz = 1.88\pm0.30\stat\pm0.15\syst$
and $\muttz = 0.91\pm0.04\stat\pm0.03\syst$, also resulting in a significance of more than five \SD for \twz production. 
The measured \ttz signal strength is compatible within two \SD with the most recent theoretical prediction 
at a$\text{N}^3$LO QCD and NLO EW~\cite{Kidonakis:2024lht}. It is compatible with previous results on \ttz 
production~\cite{CMS:2019too, CMS:2024mke, ATLAS:2023eld} when considering the 12\% increase in the 
predicted cross section compared to that used in these measurements. 
The correlation between \mutwz and \muttz is reduced compared to 
the previous CMS analysis~\cite{twzevidence} because of the improved ML algorithm and the separation of the 
\SRa into three categories. The combined effect is a better constraint on \mutwz 
largely independent of the value of \muttz.

In summary, a measurement of \twz production in proton-proton (\pp) collisions has been presented 
in events with three and four charged leptons. The analysis is performed using data recorded
with the CMS detector at center-of-mass energies of 13 and 13.6\TeV,
corresponding to a total integrated luminosity of 200\fbinv. 
Advanced machine-learning algorithms are employed to significantly 
improve the sensitivity of the analysis compared to an earlier analysis. 
Analyzing the same data, an expected significance of \sexpTwo standard deviations (\SD\unskip) is achieved, 
compared to 1.4 \SD obtained previously. Combining the 13 and 13.6\TeV data, a significance 
of \sobs{}\SD is observed.
The measured signal strength $\mutwz=\SSAll$ agrees with the standard model prediction within 2.3 \SD 
The measured cross sections are \XSTwo and \XSThree for $\sqrt{s}=13$ and 13.6\TeV, respectively. 
This is the first observation of \twz production in \pp collisions. 
This measurement represents a significant step forward in our understanding of rare processes 
involving the top quark and electroweak bosons. 
It opens up the possibility for future studies of the top quark electroweak interactions 
in the context of the standard model and its extensions.

\begin{acknowledgments}
We congratulate our colleagues in the CERN accelerator departments for the excellent performance of the LHC and thank the technical and administrative staffs at CERN and at other CMS institutes for their contributions to the success of the CMS effort. In addition, we gratefully acknowledge the computing centers and personnel of the Worldwide LHC Computing Grid and other centers for delivering so effectively the computing infrastructure essential to our analyses. Finally, we acknowledge the enduring support for the construction and operation of the LHC, the CMS detector, and the supporting computing infrastructure provided by the following funding agencies: SC (Armenia), BMBWF and FWF (Austria); FNRS and FWO (Belgium); CNPq, CAPES, FAPERJ, FAPERGS, and FAPESP (Brazil); MES and BNSF (Bulgaria); CERN; CAS, MoST, and NSFC (China); MINCIENCIAS (Colombia); MSES and CSF (Croatia); RIF (Cyprus); SENESCYT (Ecuador); ERC PRG, TARISTU24-TK10 and MoER TK202 (Estonia); Academy of Finland, MEC, and HIP (Finland); CEA and CNRS/IN2P3 (France); SRNSF (Georgia); BMFTR, DFG, and HGF (Germany); GSRI (Greece); NKFIH (Hungary); DAE and DST (India); IPM (Iran); SFI (Ireland); INFN (Italy); MSIT and NRF (Republic of Korea); MES (Latvia); LMTLT (Lithuania); MOE and UM (Malaysia); BUAP, CINVESTAV, CONACYT, LNS, SEP, and UASLP-FAI (Mexico); MOS (Montenegro); MBIE (New Zealand); PAEC (Pakistan); MES, NSC, and NAWA (Poland); FCT (Portugal);  MESTD (Serbia); MICIU/AEI and PCTI (Spain); MOSTR (Sri Lanka); Swiss Funding Agencies (Switzerland); MST (Taipei); MHESI (Thailand); TUBITAK and TENMAK (T\"{u}rkiye); NASU (Ukraine); STFC (United Kingdom); DOE and NSF (USA).
\end{acknowledgments}

\bibliography{auto_generated} 

\providecommand{\href}[2]{#2}\begingroup\raggedright\begin{thebibliography}{10}%
\makeatletter
\providecommand{\hrefCMSnoop }[0]{\@secondoftwo}%
\makeatother
\providecommand{\doi}{\texttt{doi:}\begingroup \urlstyle{tt}\Url}

\bibitem{Bertone:2004pz}
\hrefCMSnoop {}{G.~Bertone, D.~Hooper, and J.~Silk, ``Particle dark matter:
  {E}vidence, candidates and constraints'',} \textit{ Phys. Rept.} \textbf{
  405} (2005) 279,
  \href{http://dx.doi.org/10.1016/j.physrep.2004.08.031}{\doi{10.1016/j.physrep.2004.08.031}},
  \href{http://www.arXiv.org/abs/hep-ph/0404175}{\texttt{arXiv:hep-ph/0404175}}.

\bibitem{Canetti:2012zc}
\hrefCMSnoop {}{L.~Canetti, M.~Drewes, and M.~Shaposhnikov, ``Matter and
  antimatter in the {U}niverse'',} \textit{ New J. Phys.} \textbf{ 14} (2012)
  095012,
  \href{http://dx.doi.org/10.1088/1367-2630/14/9/095012}{\doi{10.1088/1367-2630/14/9/095012}},
  \href{http://www.arXiv.org/abs/1204.4186}{\texttt{arXiv:1204.4186}}.

\bibitem{Deppisch:2015qwa}
\hrefCMSnoop {}{F.~F. Deppisch, P.~S. Bhupal~Dev, and A.~Pilaftsis, ``Neutrinos
  and collider physics'',} \textit{ New J. Phys.} \textbf{ 17} (2015) 075019,
  \href{http://dx.doi.org/10.1088/1367-2630/17/7/075019}{\doi{10.1088/1367-2630/17/7/075019}},
  \href{http://www.arXiv.org/abs/1502.06541}{\texttt{arXiv:1502.06541}}.

\bibitem{Addazi:2021xuf}
\hrefCMSnoop {}{A.~Addazi { et~al.}, ``Quantum gravity phenomenology at the
  dawn of the multi-messenger era{\textemdash}{A} review'',} \textit{ Prog.
  Part. Nucl. Phys.} \textbf{ 125} (2022) 103948,
  \href{http://dx.doi.org/10.1016/j.ppnp.2022.103948}{\doi{10.1016/j.ppnp.2022.103948}},
  \href{http://www.arXiv.org/abs/2111.05659}{\texttt{arXiv:2111.05659}}.

\bibitem{Rappoccio:2018qxp}
\hrefCMSnoop {}{S.~Rappoccio, ``The experimental status of direct searches for
  exotic physics beyond the standard model at the {L}arge {H}adron
  {C}ollider'',} \textit{ Rev. Phys.} \textbf{ 4} (2019) 100027,
  \href{http://dx.doi.org/10.1016/j.revip.2018.100027}{\doi{10.1016/j.revip.2018.100027}},
  \href{http://www.arXiv.org/abs/1810.10579}{\texttt{arXiv:1810.10579}}.

\bibitem{ATLAS:2024itc}
\hrefCMSnoop {}{{ATLAS Collaboration}, ``{ATLAS} searches for additional
  scalars and exotic {H}iggs boson decays with the {LHC Run~2} dataset'',}
  \textit{ Phys. Rept.} \textbf{ 1116} (2025) 184,
  \href{http://dx.doi.org/10.1016/j.physrep.2024.09.002}{\doi{10.1016/j.physrep.2024.09.002}},
  \href{http://www.arXiv.org/abs/2405.04914}{\texttt{arXiv:2405.04914}}.

\bibitem{ATLAS:2024lda}
\hrefCMSnoop {}{{ATLAS Collaboration}, ``The quest to discover supersymmetry at
  the {ATLAS} experiment'',} \textit{ Phys. Rept.} \textbf{ 1116} (2025) 261,
  \href{http://dx.doi.org/10.1016/j.physrep.2024.09.010}{\doi{10.1016/j.physrep.2024.09.010}},
  \href{http://www.arXiv.org/abs/2403.02455}{\texttt{arXiv:2403.02455}}.

\bibitem{ATLAS:2024fdw}
\hrefCMSnoop {}{{ATLAS Collaboration}, ``Exploration at the high-energy
  frontier: {ATLAS Run~2} searches investigating the exotic jungle beyond the
  standard model'',} \textit{ Phys. Rept.} \textbf{ 1116} (2025) 301,
  \href{http://dx.doi.org/10.1016/j.physrep.2024.10.001}{\doi{10.1016/j.physrep.2024.10.001}},
  \href{http://www.arXiv.org/abs/2403.09292}{\texttt{arXiv:2403.09292}}.

\bibitem{CMS:2024phk}
\hrefCMSnoop {}{{CMS Collaboration}, ``Searches for {H}iggs boson production
  through decays of heavy resonances'',} \textit{ Phys. Rept.} \textbf{ 1115}
  (2025) 368,
  \href{http://dx.doi.org/10.1016/j.physrep.2024.09.004}{\doi{10.1016/j.physrep.2024.09.004}},
  \href{http://www.arXiv.org/abs/2403.16926}{\texttt{arXiv:2403.16926}}.

\bibitem{CMS:2024zqs}
\hrefCMSnoop {}{{CMS Collaboration}, ``Dark sector searches with the {CMS}
  experiment'',} \textit{ Phys. Rept.} \textbf{ 1115} (2025) 448,
  \href{http://dx.doi.org/10.1016/j.physrep.2024.09.013}{\doi{10.1016/j.physrep.2024.09.013}},
  \href{http://www.arXiv.org/abs/2405.13778}{\texttt{arXiv:2405.13778}}.

\bibitem{CMS:2024bni}
\hrefCMSnoop {}{{CMS Collaboration}, ``Review of searches for vector-like
  quarks, vector-like leptons, and heavy neutral leptons in
  proton{\textendash}proton collisions at {$\sqrt{s}=13$} {TeV} at the {CMS}
  experiment'',} \textit{ Phys. Rept.} \textbf{ 1115} (2025) 570,
  \href{http://dx.doi.org/10.1016/j.physrep.2024.09.012}{\doi{10.1016/j.physrep.2024.09.012}},
  \href{http://www.arXiv.org/abs/2405.17605}{\texttt{arXiv:2405.17605}}.

\bibitem{CMS:2024zhe}
\hrefCMSnoop {}{{CMS Collaboration}, ``Enriching the physics program of the
  {CMS} experiment via data scouting and data parking'',} \textit{ Phys. Rept.}
  \textbf{ 1115} (2025) 678,
  \href{http://dx.doi.org/10.1016/j.physrep.2024.09.006}{\doi{10.1016/j.physrep.2024.09.006}},
  \href{http://www.arXiv.org/abs/2403.16134}{\texttt{arXiv:2403.16134}}.

\bibitem{CMS:2024gzs}
\hrefCMSnoop {}{{CMS Collaboration}, ``Stairway to discovery: {A} report on the
  {CMS} programme of cross section measurements from millibarns to
  femtobarns'',} \textit{ Phys. Rept.} \textbf{ 1115} (2025) 3,
  \href{http://dx.doi.org/10.1016/j.physrep.2024.11.005}{\doi{10.1016/j.physrep.2024.11.005}},
  \href{http://www.arXiv.org/abs/2405.18661}{\texttt{arXiv:2405.18661}}.

\bibitem{London:2021lfn}
\hrefCMSnoop {}{D.~London and J.~Matias, ``{$B$} flavour anomalies: 2021
  theoretical status report'',} \textit{ Ann. Rev. Nucl. Part. Sci.} \textbf{
  72} (2022) 37,
  \href{http://dx.doi.org/10.1146/annurev-nucl-102020-090209}{\doi{10.1146/annurev-nucl-102020-090209}},
  \href{http://www.arXiv.org/abs/2110.13270}{\texttt{arXiv:2110.13270}}.

\bibitem{Maltoni:2019aot}
\hrefCMSnoop {}{F.~Maltoni, L.~Mantani, and K.~Mimasu, ``Top-quark electroweak
  interactions at high energy'',} \textit{ JHEP} \textbf{ 10} (2019) 004,
  \href{http://dx.doi.org/10.1007/JHEP10(2019)004}{\doi{10.1007/JHEP10(2019)004}},
  \href{http://www.arXiv.org/abs/1904.05637}{\texttt{arXiv:1904.05637}}.

\bibitem{Alwall:2014hca}
J.~Alwall\hrefCMSnoop {}{ { et~al.}, ``The automated computation of tree-level
  and next-to-leading order differential cross sections, and their matching to
  parton shower simulations'',} \textit{ JHEP} \textbf{ 07} (2014) 079,
  \href{http://dx.doi.org/10.1007/JHEP07(2014)079}{\doi{10.1007/JHEP07(2014)079}},
  \href{http://www.arXiv.org/abs/1405.0301}{\texttt{arXiv:1405.0301}}.

\bibitem{Frederix2012}
\hrefCMSnoop {}{R.~Frederix and S.~Frixione, ``Merging meets matching in
  {\MCATNLO}'',} \textit{ JHEP} \textbf{ 12} (2012) 061,
  \href{http://dx.doi.org/10.1007/JHEP12(2012)061}{\doi{10.1007/JHEP12(2012)061}},
  \href{http://www.arXiv.org/abs/1209.6215}{\texttt{arXiv:1209.6215}}.

\bibitem{Faham:2021zet}
\hrefCMSnoop {}{H.~El~Faham, F.~Maltoni, K.~Mimasu, and M.~Zaro, ``Single top
  production in association with a ${\PW\PZ}$ pair at the {LHC} in the
  {SMEFT}'',} \textit{ JHEP} \textbf{ 01} (2022) 100,
  \href{http://dx.doi.org/10.1007/JHEP01(2022)100}{\doi{10.1007/JHEP01(2022)100}},
  \href{http://www.arXiv.org/abs/2111.03080}{\texttt{arXiv:2111.03080}}.

\bibitem{Keaveney:2021dfa}
\hrefCMSnoop {}{J.~Keaveney, ``Constraining the {SMEFT} with a differential
  cross section measurement of ${\PQt\PW\PZ}$ production at the {HL-LHC}'',}
  \textit{ Phys. Rev. D} \textbf{ 107} (2023) 036021,
  \href{http://dx.doi.org/10.1103/PhysRevD.107.036021}{\doi{10.1103/PhysRevD.107.036021}},
  \href{http://www.arXiv.org/abs/2107.01053}{\texttt{arXiv:2107.01053}}.

\bibitem{Belvedere:2024wzg}
\hrefCMSnoop {}{A.~Belvedere, C.~Englert, R.~Kogler, and M.~Spannowsky,
  ``Dispelling the {$\sqrt{\mathcal {L}}$} myth for the {High-Luminosity
  LHC}'',} \textit{ Eur. Phys. J. C} \textbf{ 84} (2024) 715,
  \href{http://dx.doi.org/10.1140/epjc/s10052-024-13032-w}{\doi{10.1140/epjc/s10052-024-13032-w}},
  \href{http://www.arXiv.org/abs/2402.07985}{\texttt{arXiv:2402.07985}}.

\bibitem{twzevidence}
\hrefCMSnoop {}{{CMS Collaboration}, ``Evidence for ${\PQt\PW\PZ}$ production
  in proton-proton collisions at $\sqrt{s}={13\TeV}$ in multilepton final
  states'',} \textit{ Phys. Lett. B} \textbf{ 855} (2024) 138815,
  \href{http://dx.doi.org/10.1016/j.physletb.2024.138815}{\doi{10.1016/j.physletb.2024.138815}},
  \href{http://www.arXiv.org/abs/2312.11668}{\texttt{arXiv:2312.11668}}.

\bibitem{hepdata}
\hrefCMSnoop {}{}{HEPD}ata record for this analysis, 2025.
\newblock
  \href{http://dx.doi.org/10.17182/hepdata.158454.v1}{\doi{10.17182/hepdata.158454.v1}}.

\bibitem{CMS:2008xjf}
\hrefCMSnoop {}{{CMS Collaboration}, ``The {CMS} experiment at the {CERN}
  {LHC}'',} \textit{ JINST} \textbf{ 3} (2008) S08004,
  \href{http://dx.doi.org/10.1088/1748-0221/3/08/S08004}{\doi{10.1088/1748-0221/3/08/S08004}}.

\bibitem{CMS:2023gfb}
\hrefCMSnoop {}{{CMS Collaboration}, ``Development of the {CMS} detector for
  the {CERN} {LHC} \mbox{Run 3}'',} \textit{ JINST} \textbf{ 19} (2024) P05064,
  \href{http://dx.doi.org/10.1088/1748-0221/19/05/P05064}{\doi{10.1088/1748-0221/19/05/P05064}},
  \href{http://www.arXiv.org/abs/2309.05466}{\texttt{arXiv:2309.05466}}.

\bibitem{CMS:2020cmk}
\hrefCMSnoop {}{{CMS Collaboration}, ``Performance of the {CMS} {\Lone} trigger
  in proton-proton collisions at $\sqrt{s}={13\TeV}$'',} \textit{ JINST}
  \textbf{ 15} (2020) P10017,
  \href{http://dx.doi.org/10.1088/1748-0221/15/10/P10017}{\doi{10.1088/1748-0221/15/10/P10017}},
  \href{http://www.arXiv.org/abs/2006.10165}{\texttt{arXiv:2006.10165}}.

\bibitem{CMS:2016ngn}
\hrefCMSnoop {}{{CMS Collaboration}, ``The {CMS} trigger system'',} \textit{
  JINST} \textbf{ 12} (2017) P01020,
  \href{http://dx.doi.org/10.1088/1748-0221/12/01/P01020}{\doi{10.1088/1748-0221/12/01/P01020}},
  \href{http://www.arXiv.org/abs/1609.02366}{\texttt{arXiv:1609.02366}}.

\bibitem{CMS:2024aqx}
\hrefCMSnoop {}{{CMS Collaboration}, ``Performance of the {CMS} high-level
  trigger during {LHC} \mbox{Run 2}'',} \textit{ JINST} \textbf{ 19} (2024)
  P11021,
  \href{http://dx.doi.org/10.1088/1748-0221/19/11/P11021}{\doi{10.1088/1748-0221/19/11/P11021}},
  \href{http://www.arXiv.org/abs/2410.17038}{\texttt{arXiv:2410.17038}}.

\bibitem{CMS:2020uim}
\hrefCMSnoop {}{{CMS Collaboration}, ``Electron and photon reconstruction and
  identification with the {CMS} experiment at the {CERN} {LHC}'',} \textit{
  JINST} \textbf{ 16} (2021) P05014,
  \href{http://dx.doi.org/10.1088/1748-0221/16/05/P05014}{\doi{10.1088/1748-0221/16/05/P05014}},
  \href{http://www.arXiv.org/abs/2012.06888}{\texttt{arXiv:2012.06888}}.

\bibitem{CMS:2018rym}
\hrefCMSnoop {}{{CMS Collaboration}, ``Performance of the {CMS} muon detector
  and muon reconstruction with proton-proton collisions at
  $\sqrt{s}={13\TeV}$'',} \textit{ JINST} \textbf{ 13} (2018) P06015,
  \href{http://dx.doi.org/10.1088/1748-0221/13/06/P06015}{\doi{10.1088/1748-0221/13/06/P06015}},
  \href{http://www.arXiv.org/abs/1804.04528}{\texttt{arXiv:1804.04528}}.

\bibitem{CMS:2014pgm}
\hrefCMSnoop {}{{CMS Collaboration}, ``Description and performance of track and
  primary-vertex reconstruction with the {CMS} tracker'',} \textit{ JINST}
  \textbf{ 9} (2014) P10009,
  \href{http://dx.doi.org/10.1088/1748-0221/9/10/P10009}{\doi{10.1088/1748-0221/9/10/P10009}},
  \href{http://www.arXiv.org/abs/1405.6569}{\texttt{arXiv:1405.6569}}.

\bibitem{CMS:2017yfk}
\hrefCMSnoop {}{{CMS Collaboration}, ``Particle-flow reconstruction and global
  event description with the {CMS} detector'',} \textit{ JINST} \textbf{ 12}
  (2017) P10003,
  \href{http://dx.doi.org/10.1088/1748-0221/12/10/P10003}{\doi{10.1088/1748-0221/12/10/P10003}},
  \href{http://www.arXiv.org/abs/1706.04965}{\texttt{arXiv:1706.04965}}.

\bibitem{CMS:2018jrd}
\hrefCMSnoop {}{{CMS Collaboration}, ``Performance of reconstruction and
  identification of {\PGt} leptons decaying to hadrons and {\PGnGt} in
  ${\Pp\Pp}$ collisions at $\sqrt{s}={13\TeV}$'',} \textit{ JINST} \textbf{ 13}
  (2018) P10005,
  \href{http://dx.doi.org/10.1088/1748-0221/13/10/P10005}{\doi{10.1088/1748-0221/13/10/P10005}},
  \href{http://www.arXiv.org/abs/1809.02816}{\texttt{arXiv:1809.02816}}.

\bibitem{CMS:2016lmd}
\hrefCMSnoop {}{{CMS Collaboration}, ``Jet energy scale and resolution in the
  {CMS} experiment in ${\Pp\Pp}$ collisions at {8\TeV}'',} \textit{ JINST}
  \textbf{ 12} (2017) P02014,
  \href{http://dx.doi.org/10.1088/1748-0221/12/02/P02014}{\doi{10.1088/1748-0221/12/02/P02014}},
  \href{http://www.arXiv.org/abs/1607.03663}{\texttt{arXiv:1607.03663}}.

\bibitem{CMS:2019ctu}
\hrefCMSnoop {}{{CMS Collaboration}, ``Performance of missing transverse
  momentum reconstruction in proton-proton collisions at $\sqrt{s}={13\TeV}$
  using the {CMS} detector'',} \textit{ JINST} \textbf{ 14} (2019) P07004,
  \href{http://dx.doi.org/10.1088/1748-0221/14/07/P07004}{\doi{10.1088/1748-0221/14/07/P07004}},
  \href{http://www.arXiv.org/abs/1903.06078}{\texttt{arXiv:1903.06078}}.

\bibitem{CMS:2021ugv}
\hrefCMSnoop {}{{CMS Collaboration}, ``Inclusive and differential cross section
  measurements of single top quark production in association with a {\PZ} boson
  in proton-proton collisions at $\sqrt{s}={13\TeV}$'',} \textit{ JHEP}
  \textbf{ 02} (2022) 107,
  \href{http://dx.doi.org/10.1007/JHEP02(2022)107}{\doi{10.1007/JHEP02(2022)107}},
  \href{http://www.arXiv.org/abs/2111.02860}{\texttt{arXiv:2111.02860}}.

\bibitem{Hirschi:2015iia}
\hrefCMSnoop {}{V.~Hirschi and O.~Mattelaer, ``Automated event generation for
  loop-induced processes'',} \textit{ JHEP} \textbf{ 10} (2015) 146,
  \href{http://dx.doi.org/10.1007/JHEP10(2015)146}{\doi{10.1007/JHEP10(2015)146}},
  \href{http://www.arXiv.org/abs/1507.00020}{\texttt{arXiv:1507.00020}}.

\bibitem{Frixione:2007zp}
\hrefCMSnoop {}{S.~Frixione, E.~Laenen, P.~Motylinski, and B.~R. Webber,
  ``Angular correlations of lepton pairs from vector boson and top quark decays
  in {M}onte {C}arlo simulations'',} \textit{ JHEP} \textbf{ 04} (2007) 081,
  \href{http://dx.doi.org/10.1088/1126-6708/2007/04/081}{\doi{10.1088/1126-6708/2007/04/081}},
  \href{http://www.arXiv.org/abs/hep-ph/0702198}{\texttt{arXiv:hep-ph/0702198}}.

\bibitem{Artoisenet:2012st}
\hrefCMSnoop {}{P.~Artoisenet, R.~Frederix, O.~Mattelaer, and R.~Rietkerk,
  ``Automatic spin-entangled decays of heavy resonances in {Monte Carlo}
  simulations'',} \textit{ JHEP} \textbf{ 03} (2013) 015,
  \href{http://dx.doi.org/10.1007/JHEP03(2013)015}{\doi{10.1007/JHEP03(2013)015}},
\href{http://www.arXiv.org/abs/1212.3460}{\texttt{arXiv:1212.3460}}.
%%CITATION = ARXIV:1212.3460;%%.

\bibitem{Frixione:2019fxg}
S.~Frixione\hrefCMSnoop {}{ { et~al.}, ``Automated simulations beyond the
  standard model: {S}upersymmetry'',} \textit{ JHEP} \textbf{ 12} (2019) 008,
  \href{http://dx.doi.org/10.1007/JHEP12(2019)008}{\doi{10.1007/JHEP12(2019)008}},
  \href{http://www.arXiv.org/abs/1907.04898}{\texttt{arXiv:1907.04898}}.

\bibitem{Frixione:2008yi}
S.~Frixione\hrefCMSnoop {}{ { et~al.}, ``Single-top hadroproduction in
  association with a {\PW} boson'',} \textit{ JHEP} \textbf{ 07} (2008) 029,
  \href{http://dx.doi.org/10.1088/1126-6708/2008/07/029}{\doi{10.1088/1126-6708/2008/07/029}},
  \href{http://www.arXiv.org/abs/0805.3067}{\texttt{arXiv:0805.3067}}.

\bibitem{White:2009yt}
\hrefCMSnoop {}{C.~D. White, S.~Frixione, E.~Laenen, and F.~Maltoni,
  ``Isolating ${\PW\PQt}$ production at the {LHC}'',} \textit{ JHEP} \textbf{
  11} (2009) 074,
  \href{http://dx.doi.org/10.1088/1126-6708/2009/11/074}{\doi{10.1088/1126-6708/2009/11/074}},
  \href{http://www.arXiv.org/abs/0908.0631}{\texttt{arXiv:0908.0631}}.

\bibitem{Nason:2004rx}
\hrefCMSnoop {}{P.~Nason, ``A new method for combining {NLO} {QCD} with shower
  {Monte Carlo} algorithms'',} \textit{ JHEP} \textbf{ 11} (2004) 040,
  \href{http://dx.doi.org/10.1088/1126-6708/2004/11/040}{\doi{10.1088/1126-6708/2004/11/040}},
  \href{http://www.arXiv.org/abs/hep-ph/0409146}{\texttt{arXiv:hep-ph/0409146}}.

\bibitem{Frixione:2007vw}
\hrefCMSnoop {}{S.~Frixione, P.~Nason, and C.~Oleari, ``Matching {NLO} {QCD}
  computations with parton shower simulations: {T}he {\POWHEG} method'',}
  \textit{ JHEP} \textbf{ 11} (2007) 070,
  \href{http://dx.doi.org/10.1088/1126-6708/2007/11/070}{\doi{10.1088/1126-6708/2007/11/070}},
  \href{http://www.arXiv.org/abs/0709.2092}{\texttt{arXiv:0709.2092}}.

\bibitem{Frixione:2007nw}
\hrefCMSnoop {}{S.~Frixione, G.~Ridolfi, and P.~Nason, ``A positive-weight
  next-to-leading-order {Monte Carlo} for heavy flavour hadroproduction'',}
  \textit{ JHEP} \textbf{ 09} (2007) 126,
  \href{http://dx.doi.org/10.1088/1126-6708/2007/09/126}{\doi{10.1088/1126-6708/2007/09/126}},
  \href{http://www.arXiv.org/abs/0707.3088}{\texttt{arXiv:0707.3088}}.

\bibitem{Alioli:2010xd}
\hrefCMSnoop {}{S.~Alioli, P.~Nason, C.~Oleari, and E.~Re, ``A general
  framework for implementing {NLO} calculations in shower {Monte Carlo}
  programs: {T}he {\POWHEG} \textsc{box}'',} \textit{ JHEP} \textbf{ 06} (2010)
  043,
  \href{http://dx.doi.org/10.1007/JHEP06(2010)043}{\doi{10.1007/JHEP06(2010)043}},
  \href{http://www.arXiv.org/abs/1002.2581}{\texttt{arXiv:1002.2581}}.

\bibitem{Re:2010bp}
\hrefCMSnoop {}{E.~Re, ``Single-top ${\PW\PQt}$-channel production matched with
  parton showers using the {\POWHEG} method'',} \textit{ Eur. Phys. J. C}
  \textbf{ 71} (2011) 1547,
  \href{http://dx.doi.org/10.1140/epjc/s10052-011-1547-z}{\doi{10.1140/epjc/s10052-011-1547-z}},
  \href{http://www.arXiv.org/abs/1009.2450}{\texttt{arXiv:1009.2450}}.

\bibitem{Campbell:1999ah}
\hrefCMSnoop {}{J.~M. Campbell and R.~K. Ellis, ``An update on vector boson
  pair production at hadron colliders'',} \textit{ Phys. Rev. D} \textbf{ 60}
  (1999) 113006,
  \href{http://dx.doi.org/10.1103/PhysRevD.60.113006}{\doi{10.1103/PhysRevD.60.113006}},
  \href{http://www.arXiv.org/abs/hep-ph/9905386}{\texttt{arXiv:hep-ph/9905386}}.

\bibitem{Campbell:2011bn}
\hrefCMSnoop {}{J.~M. Campbell, R.~K. Ellis, and C.~Williams, ``Vector boson
  pair production at the {LHC}'',} \textit{ JHEP} \textbf{ 07} (2011) 018,
  \href{http://dx.doi.org/10.1007/JHEP07(2011)018}{\doi{10.1007/JHEP07(2011)018}},
  \href{http://www.arXiv.org/abs/1105.0020}{\texttt{arXiv:1105.0020}}.

\bibitem{Campbell:2015qma}
\hrefCMSnoop {}{J.~M. Campbell, R.~K. Ellis, and W.~T. Giele, ``A
  multi-threaded version of {\MCFM}'',} \textit{ Eur. Phys. J. C} \textbf{ 75}
  (2015) 246,
  \href{http://dx.doi.org/10.1140/epjc/s10052-015-3461-2}{\doi{10.1140/epjc/s10052-015-3461-2}},
  \href{http://www.arXiv.org/abs/1503.06182}{\texttt{arXiv:1503.06182}}.

\bibitem{NNPDF:2017mvq}
\hrefCMSnoop {}{{NNPDF} Collaboration, ``Parton distributions from
  high-precision collider data'',} \textit{ Eur. Phys. J. C} \textbf{ 77}
  (2017) 663,
  \href{http://dx.doi.org/10.1140/epjc/s10052-017-5199-5}{\doi{10.1140/epjc/s10052-017-5199-5}},
  \href{http://www.arXiv.org/abs/1706.00428}{\texttt{arXiv:1706.00428}}.

\bibitem{Sjostrand:2014zea}
T.~Sj{\"o}strand\hrefCMSnoop {}{ { et~al.}, ``An introduction to
  {\PYTHIA8.2}'',} \textit{ Comput. Phys. Commun.} \textbf{ 191} (2015) 159,
  \href{http://dx.doi.org/10.1016/j.cpc.2015.01.024}{\doi{10.1016/j.cpc.2015.01.024}},
  \href{http://www.arXiv.org/abs/1410.3012}{\texttt{arXiv:1410.3012}}.

\bibitem{Kidonakis:2024lht}
\hrefCMSnoop {}{N.~Kidonakis and C.~Foster, ``Higher-order soft-gluon
  corrections for ${\ttbar\PZ}$ cross sections'',} \textit{ Phys. Lett. B}
  \textbf{ 860} (2025) 139146,
  \href{http://dx.doi.org/10.1016/j.physletb.2024.139146}{\doi{10.1016/j.physletb.2024.139146}},
  \href{http://www.arXiv.org/abs/2410.01214}{\texttt{arXiv:2410.01214}}.

\bibitem{TuneCP5}
\hrefCMSnoop {}{{CMS Collaboration}, ``Extraction and validation of a new set
  of {CMS} {\PYTHIA8} tunes from underlying-event measurements'',} \textit{
  Eur. Phys. J. C} \textbf{ 80} (2020) 4,
  \href{http://dx.doi.org/10.1140/epjc/s10052-019-7499-4}{\doi{10.1140/epjc/s10052-019-7499-4}},
  \href{http://www.arXiv.org/abs/1903.12179}{\texttt{arXiv:1903.12179}}.

\bibitem{Alwall:2007fs}
J.~Alwall\hrefCMSnoop {}{ { et~al.}, ``Comparative study of various algorithms
  for the merging of parton showers and matrix elements in hadronic
  collisions'',} \textit{ Eur. Phys. J. C} \textbf{ 53} (2008) 473,
  \href{http://dx.doi.org/10.1140/epjc/s10052-007-0490-5}{\doi{10.1140/epjc/s10052-007-0490-5}},
  \href{http://www.arXiv.org/abs/0706.2569}{\texttt{arXiv:0706.2569}}.

\bibitem{Agostinelli:2002hh}
\hrefCMSnoop {}{{GEANT4} Collaboration, ``{\GEANTfour}---a simulation
  toolkit'',} \textit{ Nucl. Instrum. Meth. A} \textbf{ 506} (2003) 250,
  \href{http://dx.doi.org/10.1016/S0168-9002(03)01368-8}{\doi{10.1016/S0168-9002(03)01368-8}}.

\bibitem{CMS-TDR-15-02}
\hrefCMSnoop {}{{CMS Collaboration}, ``Technical proposal for the {Phase-II}
  upgrade of the {Compact Muon Solenoid}'',} CMS Technical Proposal
  CERN-LHCC-2015-010, CMS-TDR-15-02, 2015.
\newblock
  \href{http://dx.doi.org/10.17181/CERN.VU8I.D59J}{\doi{10.17181/CERN.VU8I.D59J}}.

\bibitem{CMS:2023dsu}
\hrefCMSnoop {}{{CMS Collaboration}, ``Muon identification using multivariate
  techniques in the cms experiment in proton-proton collisions at
  $\sqrt{s}={13\TeV}$'',} \textit{ JINST} \textbf{ 19} (2024) P02031,
  \href{http://dx.doi.org/10.1088/1748-0221/19/02/P02031}{\doi{10.1088/1748-0221/19/02/P02031}},
  \href{http://www.arXiv.org/abs/2310.03844}{\texttt{arXiv:2310.03844}}.

\bibitem{CMS:2023ftu}
\hrefCMSnoop {}{{CMS Collaboration}, ``Observation of four top quark production
  in proton-proton collisions at $\sqrt{s}={13\TeV}$'',} \textit{ Phys. Lett.
  B} \textbf{ 847} (2023) 138290,
  \href{http://dx.doi.org/10.1016/j.physletb.2023.138290}{\doi{10.1016/j.physletb.2023.138290}},
  \href{http://www.arXiv.org/abs/2305.13439}{\texttt{arXiv:2305.13439}}.

\bibitem{CMS:2024ild}
\hrefCMSnoop {}{{CMS Collaboration}, ``Measurement of the inclusive {$\PW\PZ$}
  production cross section in {$\Pp\Pp$} collisions at
  $\sqrt{s}={13.6\TeV}$'',} \textit{ JHEP} \textbf{ 04} (2025) 115,
  \href{http://dx.doi.org/10.1007/JHEP04(2025)115}{\doi{10.1007/JHEP04(2025)115}},
  \href{http://www.arXiv.org/abs/2412.02477}{\texttt{arXiv:2412.02477}}.

\bibitem{Cacciari:2008gp}
\hrefCMSnoop {}{M.~Cacciari, G.~P. Salam, and G.~Soyez, ``The anti-\kt jet
  clustering algorithm'',} \textit{ JHEP} \textbf{ 04} (2008) 063,
  \href{http://dx.doi.org/10.1088/1126-6708/2008/04/063}{\doi{10.1088/1126-6708/2008/04/063}},
  \href{http://www.arXiv.org/abs/0802.1189}{\texttt{arXiv:0802.1189}}.

\bibitem{Cacciari:2011ma}
\hrefCMSnoop {}{M.~Cacciari, G.~P. Salam, and G.~Soyez, ``{\FASTJET} user
  manual'',} \textit{ Eur. Phys. J. C} \textbf{ 72} (2012) 1896,
  \href{http://dx.doi.org/10.1140/epjc/s10052-012-1896-2}{\doi{10.1140/epjc/s10052-012-1896-2}},
  \href{http://www.arXiv.org/abs/1111.6097}{\texttt{arXiv:1111.6097}}.

\bibitem{Sirunyan:2017ezt}
\hrefCMSnoop {}{{CMS Collaboration}, ``Identification of heavy-flavour jets
  with the {CMS} detector in ${\Pp\Pp}$ collisions at {13\TeV}'',} \textit{
  JINST} \textbf{ 13} (2018) P05011,
  \href{http://dx.doi.org/10.1088/1748-0221/13/05/P05011}{\doi{10.1088/1748-0221/13/05/P05011}},
  \href{http://www.arXiv.org/abs/1712.07158}{\texttt{arXiv:1712.07158}}.

\bibitem{Bols:2020bkb}
E.~Bols\hrefCMSnoop {}{ { et~al.}, ``Jet flavour classification using
  {DeepJet}'',} \textit{ JINST} \textbf{ 15} (2020) P12012,
  \href{http://dx.doi.org/10.1088/1748-0221/15/12/P12012}{\doi{10.1088/1748-0221/15/12/P12012}},
  \href{http://www.arXiv.org/abs/2008.10519}{\texttt{arXiv:2008.10519}}.

\bibitem{CMS:DP-2023-005}
\href {https://cds.cern.ch/record/2854609}{{CMS Collaboration}, ``Performance
  summary of {AK4} jet {\PQb} tagging with data from proton-proton collisions
  at {13\TeV} with the {CMS} detector'',} CMS Detector Performance Note
  CMS-DP-2023-005, 2023.

\bibitem{Qu:2022mxj}
\href {https://proceedings.mlr.press/v162/qu22b.html}{H.~Qu, C.~Li, and
  S.~Qian, ``Particle transformer for jet tagging'',} in \textit{ {Proc. 39th
  International Conference on Machine Learning (ICML 2022): Baltimore MD, USA,
  July 17--23}}.
\newblock 2022.
\newblock
  \href{http://www.arXiv.org/abs/2202.03772}{\texttt{arXiv:2202.03772}}.

\bibitem{CMS-DP-2022-050}
\href {https://cds.cern.ch/record/2839920}{{CMS Collaboration}, ``Transformer
  models for heavy flavor jet identification'',} CMS Detector Performance Note
  CMS-DP-2022-050, 2022.

\bibitem{CMS-DP-2022-049}
\href {https://cds.cern.ch/record/2839919}{{CMS Collaboration}, ``Adversarial
  training for {\PQb}-tagging algorithms in {CMS}'',} CMS Detector Performance
  Note CMS-DP-2022-049, 2022.

\bibitem{CMS-DP-2024-024}
\href {https://cds.cern.ch/record/2898463}{{CMS Collaboration}, ``\mbox{Run 3}
  commissioning results of heavy-flavor jet tagging at $\sqrt{s}={13.6\TeV}$
  with {CMS} data using a modern framework for data processing'',} CMS Detector
  Performance Note CMS-DP-2022-024, 2024.

\bibitem{CMS-DP-2024-025}
\href {https://cds.cern.ch/record/2898464}{{CMS Collaboration}, ``Performance
  summary of {AK4} jet {\PQb} tagging with data from 2022 proton-proton
  collisions at {13.6\TeV} with the {CMS} detector'',} CMS Detector Performance
  Note CMS-DP-2022-025, 2024.

\bibitem{ParticleDataGroup:2024cfk}
\hrefCMSnoop {}{{Particle Data Group}, ``Review of particle physics'',}
  \textit{ Phys. Rev. D} \textbf{ 110} (2024) 030001,
  \href{http://dx.doi.org/10.1103/PhysRevD.110.030001}{\doi{10.1103/PhysRevD.110.030001}}.

\bibitem{CMS:2024onh}
\hrefCMSnoop {}{{CMS Collaboration}, ``The {CMS} statistical analysis and
  combination tool: \textsc{combine}'',} \textit{ Comput. Softw. Big Sci.}
  \textbf{ 8} (2024) 19,
  \href{http://dx.doi.org/10.1007/s41781-024-00121-4}{\doi{10.1007/s41781-024-00121-4}},
  \href{http://www.arXiv.org/abs/2404.06614}{\texttt{arXiv:2404.06614}}.

\bibitem{CMS:2020ebo}
\hrefCMSnoop {}{{CMS Collaboration}, ``Pileup mitigation at {CMS} in 13 {TeV}
  data'',} \textit{ JINST} \textbf{ 15} (2020) P09018,
  \href{http://dx.doi.org/10.1088/1748-0221/15/09/P09018}{\doi{10.1088/1748-0221/15/09/P09018}},
  \href{http://www.arXiv.org/abs/2003.00503}{\texttt{arXiv:2003.00503}}.

\bibitem{CMS-LUM-17-003}
\hrefCMSnoop {}{{CMS Collaboration}, ``Precision luminosity measurement in
  proton-proton collisions at $\sqrt{s}={13\TeV}$ in 2015 and 2016 at {CMS}'',}
  \textit{ Eur. Phys. J. C} \textbf{ 81} (2021) 800,
  \href{http://dx.doi.org/10.1140/epjc/s10052-021-09538-2}{\doi{10.1140/epjc/s10052-021-09538-2}},
  \href{http://www.arXiv.org/abs/2104.01927}{\texttt{arXiv:2104.01927}}.

\bibitem{CMS-PAS-LUM-17-004}
\href {https://cds.cern.ch/record/2621960}{{CMS Collaboration}, ``{CMS}
  luminosity measurement for the 2017 data-taking period at
  $\sqrt{s}={13\TeV}$'',} CMS Physics Analysis Summary CMS-PAS-LUM-17-004,
  2018.

\bibitem{CMS-PAS-LUM-18-002}
\href {https://cds.cern.ch/record/2676164}{{CMS Collaboration}, ``{CMS}
  luminosity measurement for the 2018 data-taking period at
  $\sqrt{s}={13\TeV}$'',} CMS Physics Analysis Summary CMS-PAS-LUM-18-002,
  2019.

\bibitem{CMS-PAS-LUM-22-001}
\href {https://cds.cern.ch/record/2890833}{{CMS Collaboration}, ``Luminosity
  measurement in proton-proton collisions at {13.6\TeV} in 2022 at {CMS}'',}
  CMS Physics Analysis Summary CMS-PAS-LUM-22-001, 2024.

\bibitem{CMS-DP-2024-068}
\href {https://cds.cern.ch/record/2904808}{{CMS Collaboration}, ``Measurement
  of the offline integrated luminosity for the {CMS} proton-proton collision
  dataset recorded in 2023'',} CMS Detector Performance Note CMS-DP-2024-068,
  2024.

\bibitem{CMS:2024mke}
\hrefCMSnoop {}{{CMS Collaboration}, ``Measurements of inclusive and
  differential cross sections for top quark production in association with a
  {\PZ} boson in proton-proton collisions at $\sqrt{s}={13\TeV}$'',} \textit{
  JHEP} \textbf{ 02} (2025) 177,
  \href{http://dx.doi.org/10.1007/JHEP02(2025)177}{\doi{10.1007/JHEP02(2025)177}},
  \href{http://www.arXiv.org/abs/2410.23475}{\texttt{arXiv:2410.23475}}.

\bibitem{ATLAS:2023eld}
\hrefCMSnoop {}{{ATLAS Collaboration}, ``Inclusive and differential
  cross-section measurements of ${\ttbar\PZ}$ production in ${\Pp\Pp}$
  collisions at $\sqrt{s}={13\TeV}$ with the {ATLAS} detector, including {EFT}
  and spin-correlation interpretations'',} \textit{ JHEP} \textbf{ 07} (2024)
  163,
  \href{http://dx.doi.org/10.1007/JHEP07(2024)163}{\doi{10.1007/JHEP07(2024)163}},
  \href{http://www.arXiv.org/abs/2312.04450}{\texttt{arXiv:2312.04450}}.

\bibitem{CMS:2021icx}
\hrefCMSnoop {}{{CMS Collaboration}, ``Measurement of the inclusive and
  differential {$\PW\PZ$} production cross sections, polarization angles, and
  triple gauge couplings in {$\Pp\Pp$} collisions at $\sqrt{s}={13\TeV}$'',}
  \textit{ JHEP} \textbf{ 07} (2022) 032,
  \href{http://dx.doi.org/10.1007/JHEP07(2022)032}{\doi{10.1007/JHEP07(2022)032}},
  \href{http://www.arXiv.org/abs/2110.11231}{\texttt{arXiv:2110.11231}}.

\bibitem{CMS-SMP-19-014}
\hrefCMSnoop {}{{CMS Collaboration}, ``Observation of the production of three
  massive gauge bosons at {$\sqrt{s}=13\TeV$}'',} \textit{ Phys. Rev. Lett.}
  \textbf{ 125} (2020) 151802,
  \href{http://dx.doi.org/10.1103/PhysRevLett.125.151802}{\doi{10.1103/PhysRevLett.125.151802}},
  \href{http://www.arXiv.org/abs/2006.11191}{\texttt{arXiv:2006.11191}}.

\bibitem{ATLAS-WGG}
\hrefCMSnoop {}{{ATLAS Collaboration}, ``Observation of {$\PW\PGg\PGg$}
  triboson production in proton-proton collisions at {$\sqrt{s}=13\TeV$} with
  the {ATLAS} detector'',} \textit{ Phys. Lett. B} \textbf{ 848} (2024) 138400,
  \href{http://dx.doi.org/10.1016/j.physletb.2023.138400}{\doi{10.1016/j.physletb.2023.138400}},
  \href{http://www.arXiv.org/abs/2308.03041}{\texttt{arXiv:2308.03041}}.

\bibitem{PhysRevLett.132.021802}
\hrefCMSnoop {}{{ATLAS Collaboration}, ``Observation of {$\PW\PZ\PGg$}
  production in {$\Pp\Pp$} collisions at $\sqrt{s}={13\TeV}$ with the {ATLAS}
  detector'',} \textit{ Phys. Rev. Lett.} \textbf{ 132} (2024) 021802,
  \href{http://dx.doi.org/10.1103/PhysRevLett.132.021802}{\doi{10.1103/PhysRevLett.132.021802}},
\href{http://www.arXiv.org/abs/2305.16994}{\texttt{arXiv:2305.16994}}.
%%CITATION = ARXIV:2305.16994;%%.

\bibitem{CMS-SMP-22-006}
\hrefCMSnoop {}{{CMS Collaboration}, ``Observation of {$\PW\PW\PGg$} production
  and search for ${\PH\PGg}$ production in proton-proton collisions at
  {$\sqrt{s}=13\TeV$}'',} \textit{ Phys. Rev. Lett.} \textbf{ 132} (2024)
  121901,
  \href{http://dx.doi.org/10.1103/PhysRevLett.132.121901}{\doi{10.1103/PhysRevLett.132.121901}},
  \href{http://www.arXiv.org/abs/2310.05164}{\texttt{arXiv:2310.05164}}.

\bibitem{Czakon:2011xx}
\hrefCMSnoop {}{M.~Czakon and A.~Mitov, ``Top++: {A} program for the
  calculation of the top-pair cross-section at hadron colliders'',} \textit{
  Comput. Phys. Commun.} \textbf{ 185} (2014) 2930,
  \href{http://dx.doi.org/10.1016/j.cpc.2014.06.021}{\doi{10.1016/j.cpc.2014.06.021}},
  \href{http://www.arXiv.org/abs/1112.5675}{\texttt{arXiv:1112.5675}}.

\bibitem{CMS:2022ilp}
\hrefCMSnoop {}{{CMS Collaboration}, ``Measurement of differential cross
  sections for the production of a {\PZ} boson in association with jets in
  proton-proton collisions at {$\sqrt{s} = 13$\,TeV}'',} \textit{ Phys. Rev. D}
  \textbf{ 108} (2023) 052004,
  \href{http://dx.doi.org/10.1103/PhysRevD.108.052004}{\doi{10.1103/PhysRevD.108.052004}},
  \href{http://www.arXiv.org/abs/2205.02872}{\texttt{arXiv:2205.02872}}.

\bibitem{CMS:2022tkv}
\hrefCMSnoop {}{{CMS Collaboration}, ``Measurement of the cross section of top
  quark-antiquark pair production in association with a {\PW} boson in
  proton-proton collisions at {$\sqrt{s}=13\TeV$}'',} \textit{ JHEP} \textbf{
  07} (2023) 219,
  \href{http://dx.doi.org/10.1007/JHEP07(2023)219}{\doi{10.1007/JHEP07(2023)219}},
  \href{http://www.arXiv.org/abs/2208.06485}{\texttt{arXiv:2208.06485}}.

\bibitem{CMS:2022ytw}
\hrefCMSnoop {}{{CMS Collaboration}, ``Measurement of inclusive and
  differential cross sections for single top quark production in association
  with a {\PW} boson in proton-proton collisions at $\sqrt{s} = {13\TeV}$'',}
  \textit{ JHEP} \textbf{ 07} (2023) 046,
  \href{http://dx.doi.org/10.1007/JHEP07(2023)046}{\doi{10.1007/JHEP07(2023)046}},
  \href{http://www.arXiv.org/abs/2208.00924}{\texttt{arXiv:2208.00924}}.

\bibitem{CMS:2022dwd}
\hrefCMSnoop {}{{CMS Collaboration}, ``A portrait of the {H}iggs boson by the
  {CMS} experiment ten years after the discovery'',} \textit{ Nature} \textbf{
  607} (2022) 60,
  \href{http://dx.doi.org/10.1038/s41586-022-04892-x}{\doi{10.1038/s41586-022-04892-x}},
  \href{http://www.arXiv.org/abs/2207.00043}{\texttt{arXiv:2207.00043}}.
  [Erratum: \DOI{10.1038/s41586-023-06164-8}].

\bibitem{Barlow:1993dm}
\hrefCMSnoop {}{R.~Barlow and C.~Beeston, ``Fitting using finite {Monte Carlo}
  samples'',} \textit{ Comput. Phys. Commun.} \textbf{ 77} (1993) 219,
  \href{http://dx.doi.org/10.1016/0010-4655(93)90005-W}{\doi{10.1016/0010-4655(93)90005-W}}.

\bibitem{CMS:2019too}
\hrefCMSnoop {}{{CMS Collaboration}, ``Measurement of top quark pair production
  in association with a {\PZ} boson in proton-proton collisions at
  {$\sqrt{s}=13$\,TeV}'',} \textit{ JHEP} \textbf{ 03} (2020) 056,
  \href{http://dx.doi.org/10.1007/JHEP03(2020)056}{\doi{10.1007/JHEP03(2020)056}},
  \href{http://www.arXiv.org/abs/1907.11270}{\texttt{arXiv:1907.11270}}.

\bibitem{Bertolini:2014bba}
\hrefCMSnoop {}{D.~Bertolini, P.~Harris, M.~Low, and N.~Tran, ``Pileup per
  particle identification'',} \textit{ JHEP} \textbf{ 10} (2014) 59,
  \href{http://dx.doi.org/10.1007/JHEP10(2014)059}{\doi{10.1007/JHEP10(2014)059}},
  \href{http://www.arXiv.org/abs/1407.6013}{\texttt{arXiv:1407.6013}}.

\bibitem{metPF}
\hrefCMSnoop {}{{CMS Collaboration}, ``Performance of the {CMS} missing
  transverse momentum reconstruction in ${\Pp\Pp}$ data at
  $\sqrt{s}={8\TeV}$'',} \textit{ JINST} \textbf{ 10} (2015) P02006,
  \href{http://dx.doi.org/10.1088/1748-0221/10/02/P02006}{\doi{10.1088/1748-0221/10/02/P02006}},
  \href{http://www.arXiv.org/abs/1411.0511}{\texttt{arXiv:1411.0511}}.

\end{thebibliography}\endgroup

\clearpage
\appendix
\numberwithin{figure}{section}
\ifthenelse{\boolean{cms@external}}{\section{}}{\section{Additional material}\label{endmatter}}

\subsection{Pileup mitigation}
\label{endmatter:pileup}

Pileup can contribute additional tracks and calorimetric energy depositions to the jet momentum. 
For data from 2016--2018, we mitigate this effect by discarding charged particles identified to be originating 
from pileup vertices and apply an offset correction to correct for remaining pileup 
contributions~\cite{CMS:2017yfk,CMS:2020ebo}.
For 2022--2023 data, we apply the pileup-per-particle identification 
algorithm~\cite{Bertolini:2014bba, CMS:2020ebo}, weighting all PF candidates 
by their probability to originate from the PV~\cite{CMS:2019ctu}.
Jet energy corrections are derived from data and simulation to bring the 
measured response of jets to that of particle level jets on average~\cite{CMS:2016lmd}.
The missing transverse momentum vector \ptvecmiss is the negative vector sum of the 
transverse momenta of all PF candidates~\cite{metPF, CMS:2019ctu}. 
Corrections to the jet energies are propagated to \ptvecmiss.

\ifthenelse{\boolean{cms@external}}
{
\begin{figure}[!bp]
\centering
\includegraphics[width=0.35\textwidth]{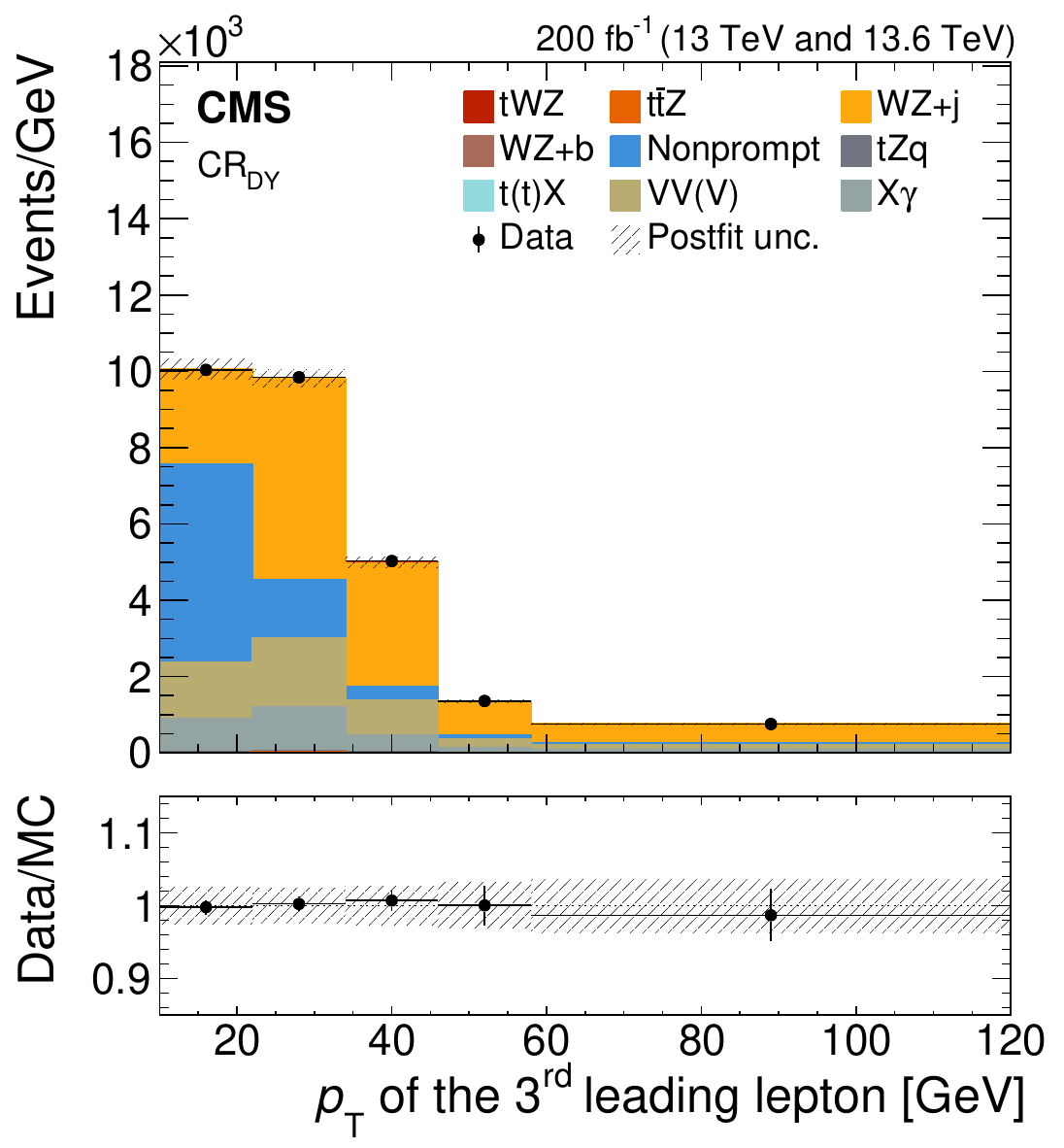} 
\hspace{3 cm}
\includegraphics[width=0.35\textwidth]{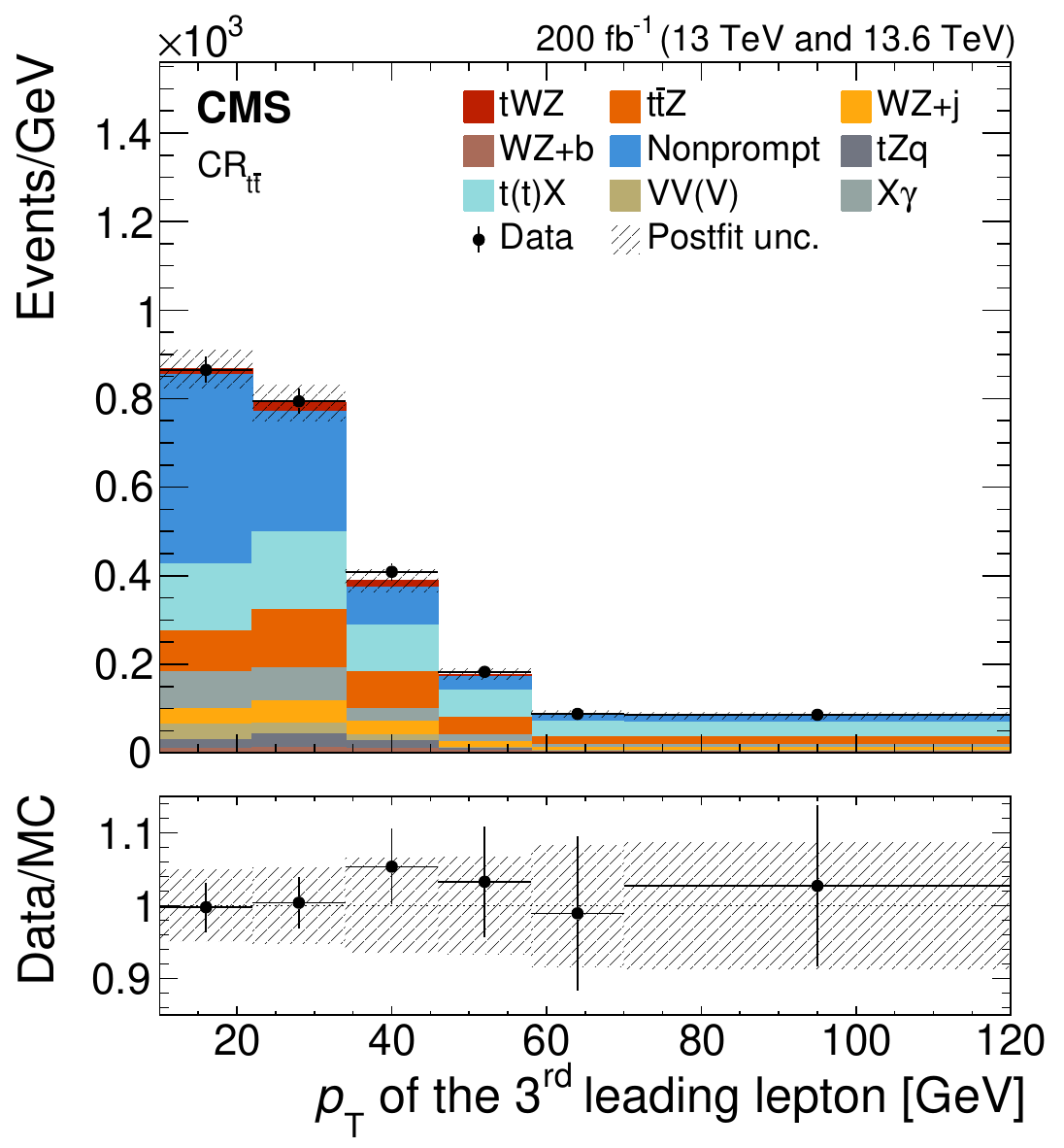}
\caption{Distribution in the \pt of the $3^{rd}$ leading lepton for the \CRDY (\cmsLeft) and \CRTT (\cmsRight) 
after a fit to data. The dashed band shows the postfit uncertainty.
The results from fits to 13 and 13.6\TeV data are combined for illustration purposes. 
The ratio of data to the predictions from MC simulations is shown below the distributions.  
\label{fig:lep3pt}
}
\end{figure}
}
{
\begin{figure}[bp]
\centering
\includegraphics[width=\cmsFigWidth]{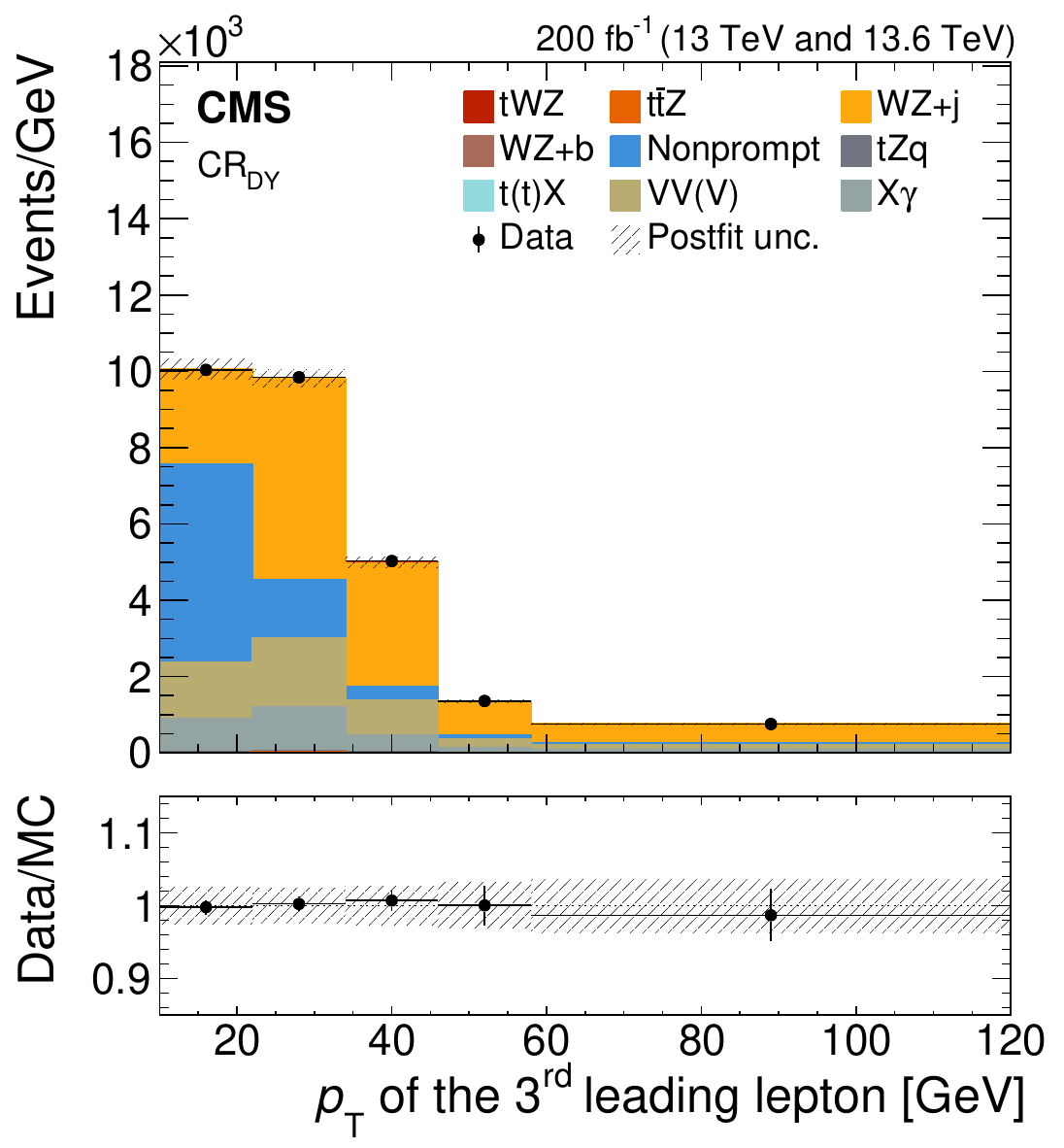}
\includegraphics[width=\cmsFigWidth]{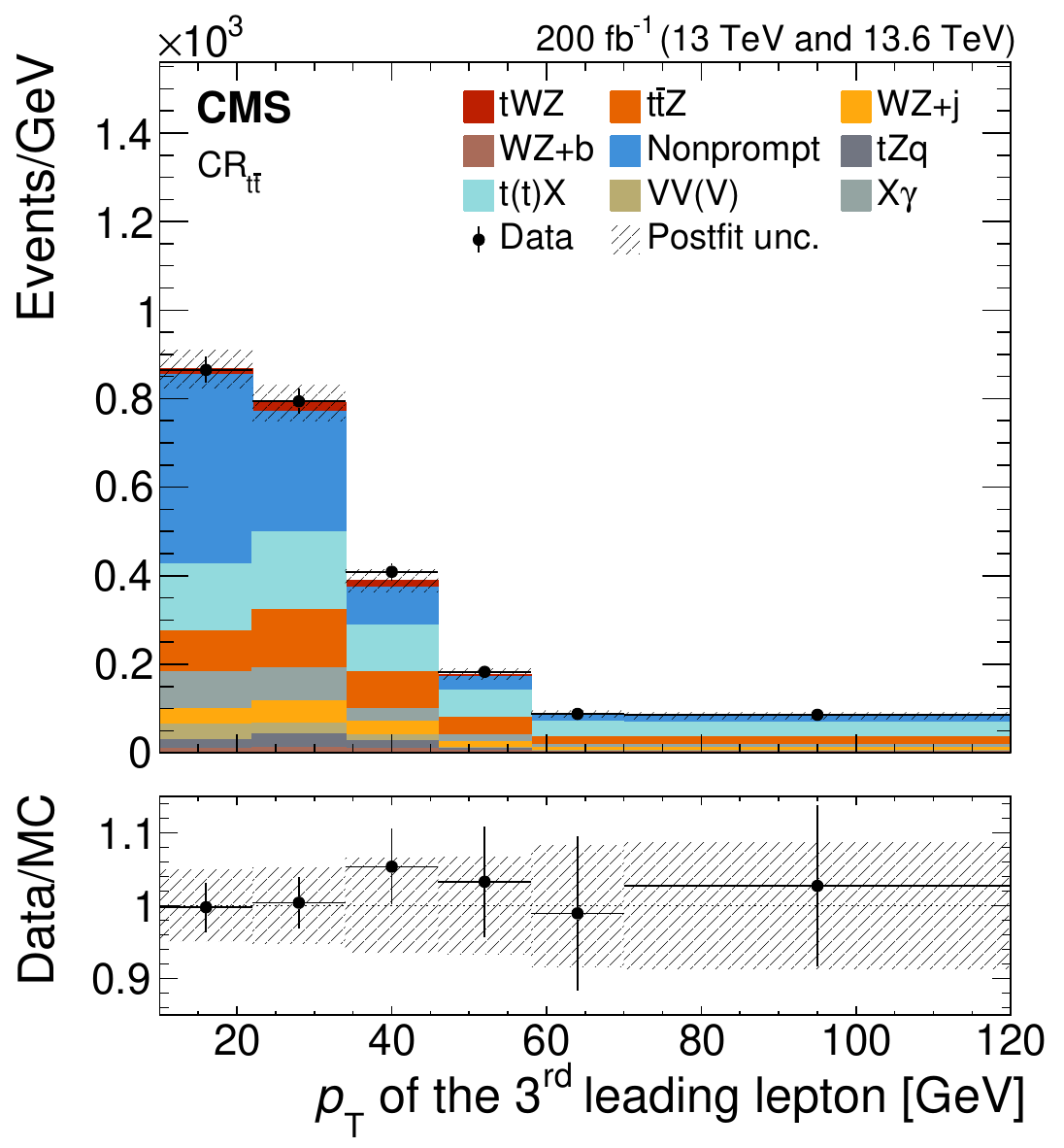}
\caption{Distribution in the \pt of the $3^{rd}$ leading lepton for the \CRDY (\cmsLeft) and \CRTT (\cmsRight) 
after a fit to data. The dashed band shows the postfit uncertainty.
The results from fits to 13 and 13.6\TeV data are combined for illustration purposes. 
The ratio of data to the predictions from MC simulations is shown below the distributions. 
\label{fig:lep3pt}
}
\end{figure}
}

\subsection{Jet multiplicity reweighting}
\label{endmatter:njet}

The fully leptonic decay of the \WZ process involves at most one jet from the matrix element 
calculation at NLO in QCD. Thus, additional jet emissions are primarily simulated by the PS, 
which may lead to mismodeling in the shape of the jet multiplicity distribution. 
This can be particularly relevant when selections that are not inclusive in the number of jets are applied.
To improve the modeling of the jet multiplicity, a simulation-to-data reweighting is performed in the \CRWZ.
Simulated contributions from processes other than \WZ are subtracted from data, and
a correction is derived as the ratio between the observed data yield and the \WZ simulation as a 
function of the jet multiplicity. A shape uncertainty is assigned to this correction by considering two alternative configurations: one where the correction factor is doubled 
and another where no correction is applied. 

\subsection{Background from nonprompt leptons}
\label{endmatter:nonprompt}

The nonprompt-lepton contribution is estimated from a fit to data using simulated templates
based on the origin of the misidentified leptons.
Events containing at least one nonprompt lepton are divided into three categories: 
nonprompt leptons from photons, \PQb jets, and light-quark and gluon jets. 
For 13.6\TeV data, nonprompt leptons from \PQb-quark, light-quark, and gluon jets are combined into a 
single category because of the smaller number of events from nonprompt leptons due to the 
improved lepton-identification algorithm. 
We introduce unconstrained parameters that describe the lepton misidentification rates as functions of \pt. 
For nonprompt leptons from photons, these parameters are considered to be normal-distributed 
around the MC predictions, with prior uncertainties of 50\%. 
Uncertainties accounting for shape differences in the trailing-lepton \pt distributions 
between nonprompt leptons of different flavors within the same category are 
considered as an additional uncertainty. We find that these have a negligible impact on the analysis. 
In the fit, the two distributions in the trailing-lepton \pt measured 
in the \CRDY and \CRTT are used. These are shown in Fig.~\ref{fig:lep3pt} after the 
fit to data. 
The nonprompt backgrounds constitute about one-third of all selected events in 
these two regions. 
The trailing-lepton \pt distributions show sensitivity to the origin of the nonprompt lepton 
and can thus constrain the parameters describing the misidentification rates. 
These are measured independently for each data-taking period. The largest correction factors 
are measured for nonprompt leptons from light-quark and gluon jets, resulting in a nonprompt-lepton background that is 
40--70\% lower than the simulation-based prediction, depending on the data-taking period.
The associated uncertainties are around 50\%, due to the small statistical power of this category. 
The correction factors associated with the nonprompt leptons from \PQb jets are closer to unity, 
where the largest correction shifts the prediction up by 40\%, with uncertainties
of about 20\%.
\ifthenelse{\boolean{cms@external}}
{
\begin{figure}[tb]
\centering
\includegraphics[width=0.33\textwidth]{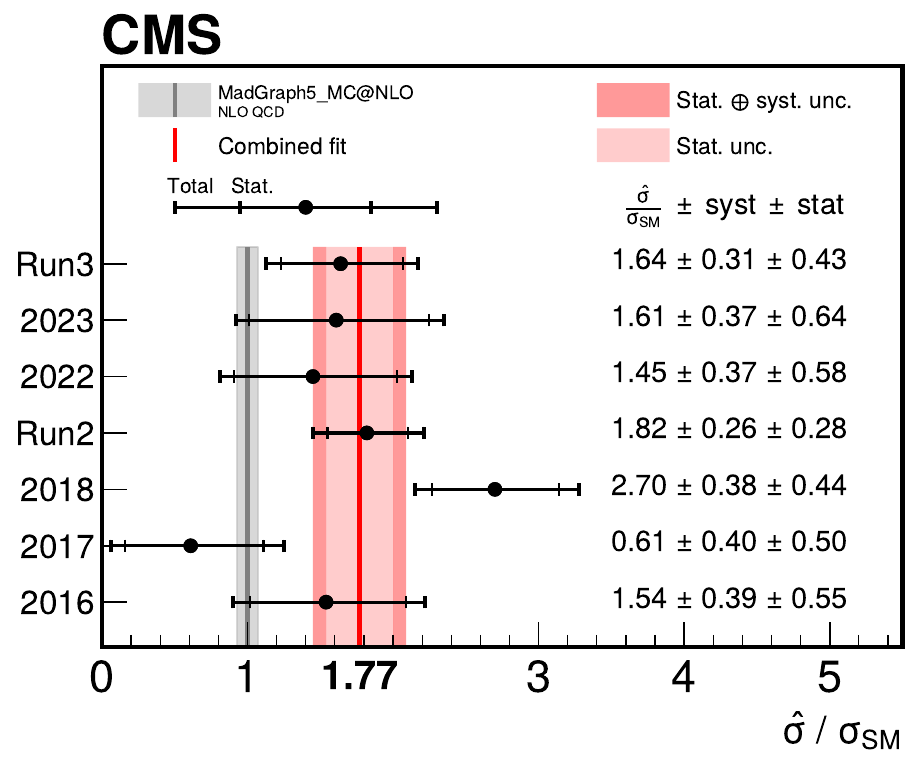}
\includegraphics[width=0.33\textwidth]{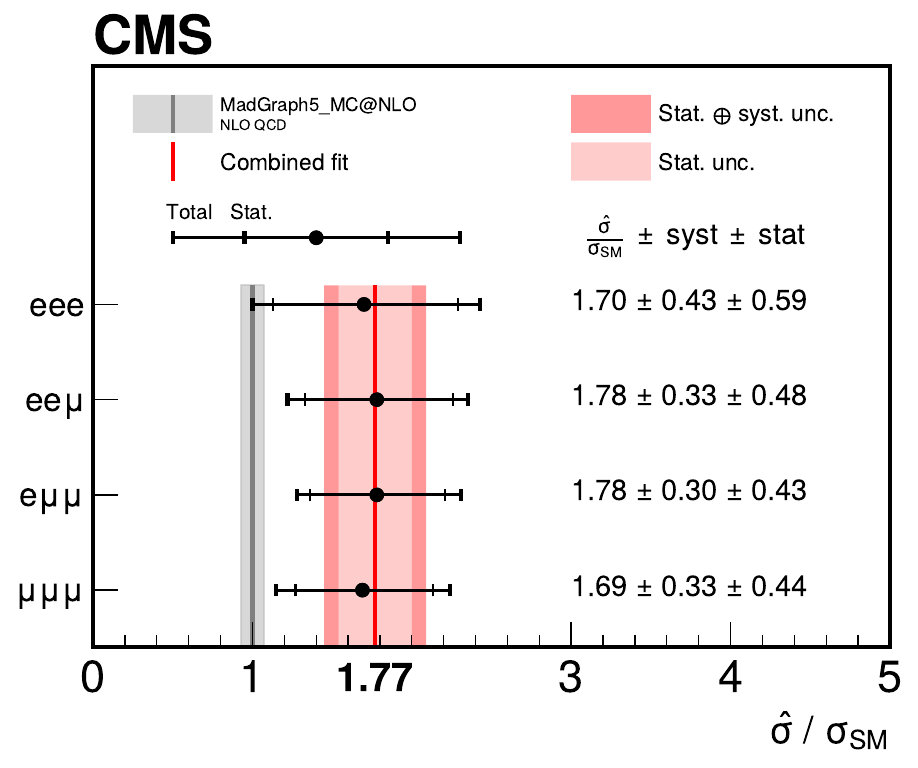}
\caption{Values of the signal strengths for \twz production obtained from 
the different data-taking periods (\cmsLeft), and from events with 
different lepton flavors (\cmsRight): three electrons ($\Pe\Pe\Pe$), two electrons and one muon ($\Pe\Pe\PGm$), one electron 
and two muons ($\Pe\PGm\PGm$), and three muons ($\PGm\PGm\PGm$). If there are more than three leptons in an event, 
the three \pt-leading leptons are used to determine the flavor. 
The combined results from 13\TeV data (Run~2) and 
from 13.6\TeV data (Run~3) are also shown. 
The result from the combined fit to 13 and 13.6\TeV data is shown as a red line 
with the corresponding uncertainty shown as a red band. The SM prediction is shown as a gray band. 
\label{fig:comp}
}
\end{figure}
}{}

\subsection{The machine-learning algorithm}
\label{endmatter:part}

The \textsc{ParT} algorithm takes as inputs two different set of variables: \textit{Particles}  
and \textit{Interactions}. 
The \textit{Particles} input corresponds to a list of individual physics objects where each 
particle is described by a set of features.  It is divided
into two categories: \textit{leptons} and \textit{jets}. The components of the four-momentum of 
each lepton and jet in every event, as well as the \ptvecmiss, are used as inputs. 
Additionally, the following set of features is considered: $\log(\pt/\GeV)$, $\log(E/\GeV)$, $\eta$, $\phi$, 
\PQb-tagging information (only for jets), and a number that encodes the flavor of the lepton.
The agreement between data and simulation has been verified
for distributions in all input variables.
The \textit{interactions} encode correlations between pairs of particles considering the
following variables: $\log(\Delta)$, $\log(\kt/\GeV)$, $\log(z)$, and $\log(m^2/\GeV^2)$.
\ifthenelse{\boolean{cms@external}}{}
{
\begin{figure}[!htbp]
\centering
\includegraphics[width=0.45\textwidth]{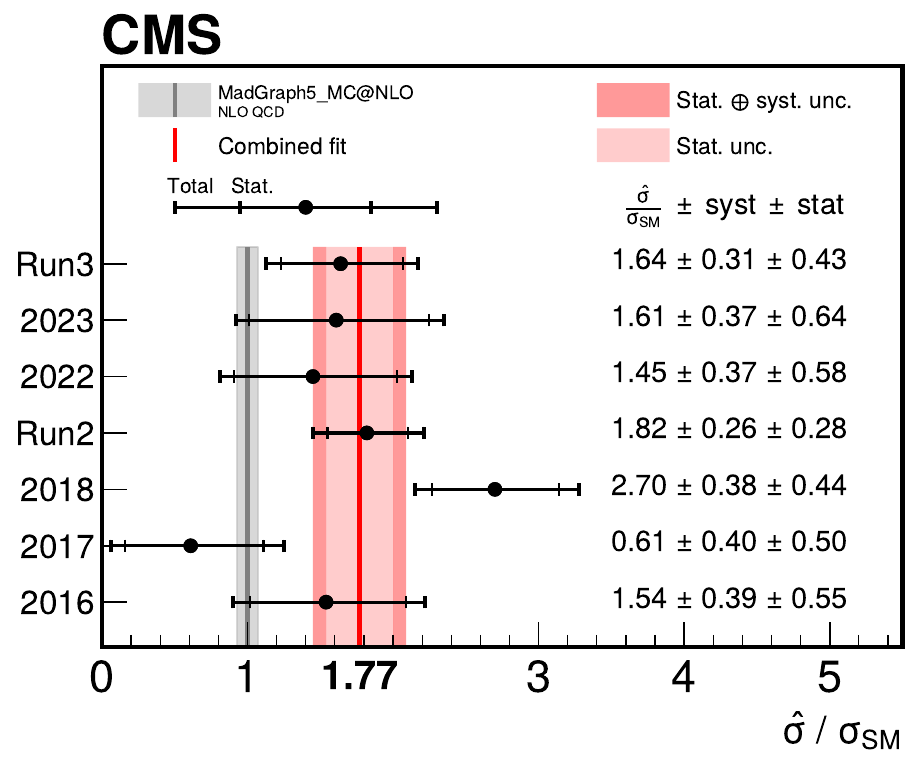}
\includegraphics[width=0.45\textwidth]{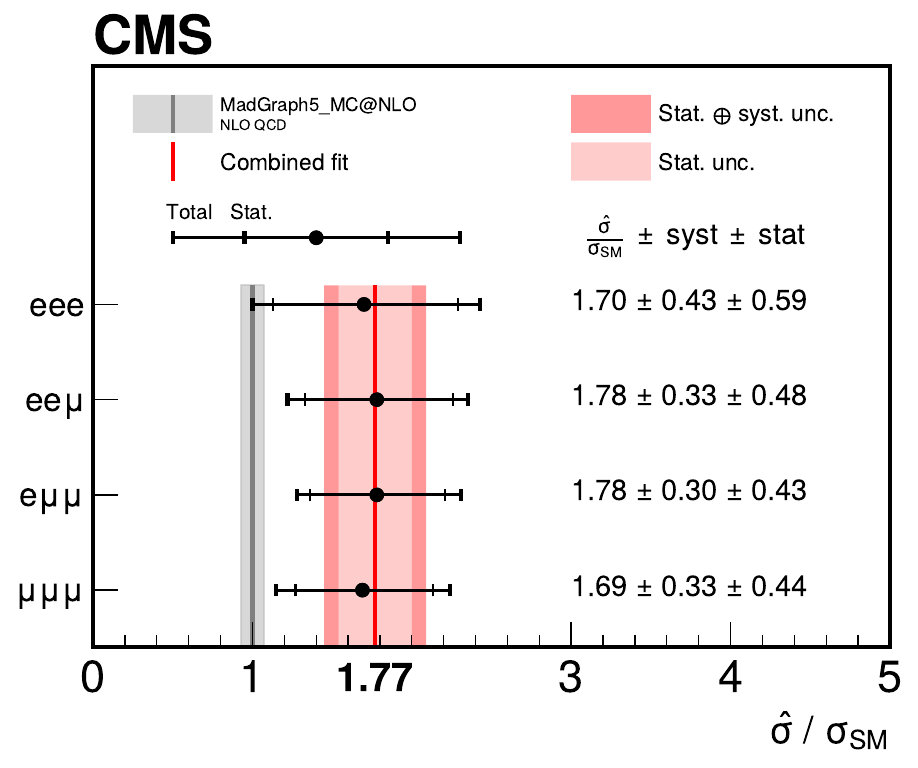}
\caption{Values of the signal strengths for \twz production obtained from 
the different data-taking periods (left), and from events with 
different lepton flavors (right): three electrons ($\Pe\Pe\Pe$), two electrons and one muon ($\Pe\Pe\PGm$), one electron 
and two muons ($\Pe\PGm\PGm$), and three muons ($\PGm\PGm\PGm$). If there are more than three leptons in an event, 
the three \pt-leading leptons are used to determine the flavor. 
The combined results from 13\TeV data (Run~2) and 
from 13.6\TeV data (Run~3) are also shown. 
The result from the combined fit to 13 and 13.6\TeV data is shown as a red line 
with the corresponding uncertainty shown as a red band. The SM prediction is shown as a gray band. 
\label{fig:comp}
}
\end{figure}
}
These are defined between two particles $a$ and $b$ as 
\begin{linenomath}
\begin{equation*}
\begin{aligned}
\Delta &= \sqrt{(y_a-y_b)^2+(\phi_a-\phi_b)^2},  \\
\kt    &=\min(p_{\mathrm{T},a},p_{\mathrm{T},b})\Delta,  \\
z       &=\min(p_{\mathrm{T},a},p_{\mathrm{T},b})/(p_{\mathrm{T},a}+p_{\mathrm{T},b}),  \\
m^2     &=(E_{a}+E_{b})^2-\abs{\mathbf{p}_{a}+\mathbf{p}_{b}}^2,
\end{aligned}
\end{equation*}
\end{linenomath}
where $y$, $\phi$, and $\mathbf{p}$ are the rapidity, azimuthal angle, and the three-momentum, respectively. 
The \textit{interactions} allow each particle to incorporate information from all other 
particles using both the particle features and their pairwise relationships, enabling the network 
to capture both local and global event characteristics. 
To avoid overtraining, two versions of the \textsc{ParT} algorithm are trained for each signal 
region and center-of-mass energy. 
An algorithm trained on one half of the simulated data is employed on the other half during the analysis. 
In this way, each algorithm is trained on and applied to separate halves of the simulated samples.
The agreement between data and simulation has also been verified in the \CRTT for the algorithms
used in the \SRa, and in the \CRZZ for the algorithms employed in the \SRb. 
These CRs have the same number of physics objects as the corresponding SRs, 
and the data distributions in the output of the algorithms show good agreement to the simulated ones. 

\subsection{Individual results}
\label{endmatter:comp}

The signal strength is also measured for the different data-taking periods and leptonic channels, defined 
by the flavor of the three leptons with the largest \pt in the event. Comparisons of 
the obtained signal strengths are shown in Fig.~\ref{fig:comp}. We find good agreement between 
the measured signal strengths in the different data-taking periods and among the signal strengths from 
different lepton flavors.

\cleardoublepage \section{The CMS Collaboration \label{app:collab}}\begin{sloppypar}\hyphenpenalty=5000\widowpenalty=500\clubpenalty=5000\cmsinstitute{Yerevan Physics Institute, Yerevan, Armenia}
{\tolerance=6000
A.~Hayrapetyan, V.~Makarenko\cmsorcid{0000-0002-8406-8605}, A.~Tumasyan\cmsAuthorMark{1}\cmsorcid{0009-0000-0684-6742}
\par}
\cmsinstitute{Institut f\"{u}r Hochenergiephysik, Vienna, Austria}
{\tolerance=6000
W.~Adam\cmsorcid{0000-0001-9099-4341}, J.W.~Andrejkovic, L.~Benato\cmsorcid{0000-0001-5135-7489}, T.~Bergauer\cmsorcid{0000-0002-5786-0293}, M.~Dragicevic\cmsorcid{0000-0003-1967-6783}, C.~Giordano, P.S.~Hussain\cmsorcid{0000-0002-4825-5278}, M.~Jeitler\cmsAuthorMark{2}\cmsorcid{0000-0002-5141-9560}, N.~Krammer\cmsorcid{0000-0002-0548-0985}, A.~Li\cmsorcid{0000-0002-4547-116X}, D.~Liko\cmsorcid{0000-0002-3380-473X}, M.~Matthewman, I.~Mikulec\cmsorcid{0000-0003-0385-2746}, J.~Schieck\cmsAuthorMark{2}\cmsorcid{0000-0002-1058-8093}, R.~Sch\"{o}fbeck\cmsAuthorMark{2}\cmsorcid{0000-0002-2332-8784}, D.~Schwarz\cmsorcid{0000-0002-3821-7331}, M.~Shooshtari\cmsorcid{0009-0004-8882-4887}, M.~Sonawane\cmsorcid{0000-0003-0510-7010}, W.~Waltenberger\cmsorcid{0000-0002-6215-7228}, C.-E.~Wulz\cmsAuthorMark{2}\cmsorcid{0000-0001-9226-5812}
% ...
\par}
\vskip\cmsinstskip
\dag:~Deceased\\
$^{1}$Also at Yerevan State University, Yerevan, Armenia\\
$^{2}$Also at TU Wien, Vienna, Austria\\
% ...
\end{sloppypar}
%%% END EDITABLE REGION %%%
% skeleton_end
\end{document}